\documentclass[10pt,twocolumn,journal]{IEEEtran}

\usepackage[mathscr]{eucal}
\usepackage[cmex10]{amsmath}
\usepackage{epsfig,epsf,psfrag}
\usepackage{amssymb,amsmath,amsthm,amsfonts,latexsym}
\usepackage{amsmath,graphicx,bm,xcolor,url}
\usepackage[caption=false]{subfig} 
\usepackage{fixltx2e}
\usepackage{array}
\usepackage{verbatim}
\usepackage{bm}
\usepackage{algorithmic, cite}
\usepackage{algorithm}
\usepackage{verbatim}
\usepackage{textcomp}
\usepackage{mathrsfs}
\usepackage{epstopdf}

\catcode`~=11 \def\UrlSpecials{\do\~{\kern -.15em\lower .7ex\hbox{~}\kern .04em}} \catcode`~=13 

\allowdisplaybreaks[1]
 
\newcommand{\nn}{\nonumber}

\newcommand{\calA}{\mathcal{A}}

\newcommand{\calC}{\mathcal{C}}

\newcommand{\calF}{\mathcal{F}}

\newcommand{\calM}{\mathcal{M}}
\newcommand{\calN}{\mathcal{N}}

\newcommand{\calP}{\mathcal{P}}

\newcommand{\calT}{\mathcal{T}}

\newcommand{\calV}{\mathcal{V}}

\newcommand{\calX}{\mathcal{X}}
\newcommand{\calY}{\mathcal{Y}}
\newcommand{\calZ}{\mathcal{Z}}

\newcommand{\bA}{\mathbf{A}}

\newcommand{\bB}{\mathbf{B}}

\newcommand{\bC}{\mathbf{C}}

\newcommand{\bI}{\mathbf{I}}

\newcommand{\bv}{\mathbf{v}}
\newcommand{\bV}{\mathbf{V}}

\newcommand{\bW}{\mathbf{W}}
\newcommand{\bx}{\mathbf{x}}
\newcommand{\bX}{\mathbf{X}}
\newcommand{\by}{\mathbf{y}}
\newcommand{\bY}{\mathbf{Y}}
\newcommand{\bz}{\mathbf{z}}
\newcommand{\bZ}{\mathbf{Z}}

\newcommand{\rmd}{\mathrm{d}}

\newcommand{\rme}{\mathrm{e}}

\newcommand{\bbE}{\mathbb{E}}

\newcommand{\bbN}{\mathbb{N}}

\newcommand{\bbR}{\mathbb{R}}
\newcommand{\bbS}{\mathbb{S}}

\newcommand{\bbV}{\mathbb{V}}

\DeclareMathAlphabet{\mathbsf}{OT1}{cmss}{bx}{n}
\DeclareMathAlphabet{\mathssf}{OT1}{cmss}{m}{sl}

\DeclareSymbolFont{bsfletters}{OT1}{cmss}{bx}{n}  
\DeclareSymbolFont{ssfletters}{OT1}{cmss}{m}{n}
\DeclareMathSymbol{\bsfGamma}{0}{bsfletters}{'000}
\DeclareMathSymbol{\ssfGamma}{0}{ssfletters}{'000}
\DeclareMathSymbol{\bsfDelta}{0}{bsfletters}{'001}
\DeclareMathSymbol{\ssfDelta}{0}{ssfletters}{'001}
\DeclareMathSymbol{\bsfTheta}{0}{bsfletters}{'002}
\DeclareMathSymbol{\ssfTheta}{0}{ssfletters}{'002}
\DeclareMathSymbol{\bsfLambda}{0}{bsfletters}{'003}
\DeclareMathSymbol{\ssfLambda}{0}{ssfletters}{'003}
\DeclareMathSymbol{\bsfXi}{0}{bsfletters}{'004}
\DeclareMathSymbol{\ssfXi}{0}{ssfletters}{'004}
\DeclareMathSymbol{\bsfPi}{0}{bsfletters}{'005}
\DeclareMathSymbol{\ssfPi}{0}{ssfletters}{'005}
\DeclareMathSymbol{\bsfSigma}{0}{bsfletters}{'006}
\DeclareMathSymbol{\ssfSigma}{0}{ssfletters}{'006}
\DeclareMathSymbol{\bsfUpsilon}{0}{bsfletters}{'007}
\DeclareMathSymbol{\ssfUpsilon}{0}{ssfletters}{'007}
\DeclareMathSymbol{\bsfPhi}{0}{bsfletters}{'010}
\DeclareMathSymbol{\ssfPhi}{0}{ssfletters}{'010}
\DeclareMathSymbol{\bsfPsi}{0}{bsfletters}{'011}
\DeclareMathSymbol{\ssfPsi}{0}{ssfletters}{'011}
\DeclareMathSymbol{\bsfOmega}{0}{bsfletters}{'012}
\DeclareMathSymbol{\ssfOmega}{0}{ssfletters}{'012}

\newcommand{\hatm}{\hat{m}}
\newcommand{\hatM}{\hat{M}}

\newcommand{\tilM}{\tilde{M}}

\newcommand{\tilX}{\tilde{X}}

\newcommand{\tilZ}{\tilde{Z}}

\newcommand{\barX}{\bar{X}}
\newcommand{\barY}{\bar{Y}}
\newcommand{\barZ}{\bar{Z}}

\newcommand{\eps}{\varepsilon}

\DeclareMathOperator*{\argmax}{arg\,max}

\DeclareMathOperator{\var}{\mathsf{Var}}

\newcommand{\bone}{\mathbf{1}}

 \def\independenT#1#2{\mathrel{\rlap{$#1#2$}\mkern5mu{#1#2}}}

\newtheorem{theorem}{Theorem}

\newtheorem{corollary}[theorem]{Corollary}
\newtheorem{definition}{Definition}

\newcommand{\qednew}{\nobreak \ifvmode \relax \else
      \ifdim\lastskip<1.5em \hskip-\lastskip
      \hskip1.5em plus0em minus0.5em \fi \nobreak
      \vrule height0.75em width0.5em depth0.25em\fi}

\DeclareMathOperator*{\plimsup}{p-lim\, sup\,}
\DeclareMathOperator*{\pliminf}{p-lim\, inf\,}

\newcommand{\underI}{\underline{I}}
\newcommand{\overI}{\overline{I}}
\newcommand{\overC}{\overline{C}}
\newcommand{\fp}{\mathfrak{q}}
\newcommand{\CDM}{C_{\mathrm{s}}^{\mathsf{DM}}}
\newcommand{\CG}{C_{\mathrm{s}}^{\mathsf{G}}}
\usepackage[ colorlinks = true,
             linkcolor = blue,
             urlcolor  = blue,
             citecolor = red,
             anchorcolor = green,
]{hyperref}
\usepackage{overpic}

\begin{document}
\flushbottom

\title{Information Spectrum Approach to Strong Converse Theorems for Degraded Wiretap Channels} 
\author{Vincent Y.~F.\ Tan$^\dagger$ $\,\,\, $  and $\,\,$ Matthieu R.~Bloch$^\ddagger$ 
\thanks{$\dagger$ V.~Y.~F. Tan is with the Department of Electrical and Computer Engineering and the Department of Mathematics, National University of Singapore, 
Singapore 117583  (Email: \url{vtan@nus.edu.sg}).   }\thanks{$^\ddagger$ M.~R.~Bloch is with the School of Electrical and Computer Engineering, Georgia Institute of Technology, Atlanta, GA 30332 and GT-CNRS UMI 2958, 2 rue Marconi, 57070 Metz, France  (Email: \url{matthieu.bloch@ece.gatech.edu}).   } \thanks{This paper was presented in part at the 52nd Annual Allerton Conference on Communication, Control, and Computing in Monticello, IL. } }


\maketitle

\begin{abstract}
We  consider  block codes for degraded wiretap channels in which the legitimate receiver decodes the message with an asymptotic error probability no larger than $\eps$ but the leakage to the eavesdropper vanishes.  For    discrete memoryless and Gaussian wiretap channels, we show that the maximum rate of transmission does not depend on $\eps \in [0,1)$, i.e., such channels possess the {\em partial strong converse} property. Furthermore, we derive sufficient conditions for the partial strong converse  property to hold for memoryless but non-stationary symmetric and degraded wiretap channels. Our proof techniques leverage   the  information spectrum method, which allows us to establish a necessary and sufficient condition for the partial strong converse to hold for general wiretap channels without any information stability assumptions. 
\end{abstract}

 
\section{Introduction}
In many modern signal processing applications~\cite{sankar},  such as credit card transactions, health informatics and device-to-device communications, a sender wishes to transmit confidential information to a  legitimate  receiver, while keeping the information private or secret from a malicious party---the  so-called eavesdropper. 
This problem is well-studied in  information-theoretic security~\cite{BlochBarros,Liang} and is known as the {\em wiretap channel} model~\cite{Wyn75} as shown in Fig.~\ref{fig:model}.
The task is to reliably communicate a message $M\in \{1,\ldots, \lceil 2^{nR} \rceil\}$ from the sender $\calX$ to the legitimate receiver  $\calY$ while keeping the eavesdropper $\calZ$ ignorant of $M$. The {\em secrecy capacity}  for the wiretap channel $W:\calX\to\calY\times\calZ$ is the supremum of all rates $R$ for which there exists a code that is {\em reliable}, i.e.,  $\calY$ can reconstruct $M$ with   probability tending to one as the blocklength $n$ tends to infinity, and {\em secure}, i.e.,  the normalized mutual information (leakage rate) of the message and the eavesdropper's signal $\frac{1}{n} I(M;Z^n)$ is arbitrarily small as $n$ grows. Wyner showed that the secrecy capacity of a degraded (i.e., $X-Y-Z$ forms a Markov chain) discrete memoryless wiretap channel  is 
\begin{equation}
\max_{P_X} \,\, I(X;Y)-I(X;Z) \quad\mbox{bits per channel use}. \label{eqn:sc}
\end{equation}
This result was generalized by Csisz\'ar and K\"orner~\cite{CsiszarKorner78} to non-degraded channels. 

In this paper, we relax the {\em reliability} condition of the wiretap code. More precisely, we allow the wiretap code to be such that the legitimate receiver    decodes the message $M$ with  an asymptotic error  probability  bounded above by $\eps \in [0,1)$. The wiretap code, however, must ensure that $M$ and $Z^n$ are asymptotically independent and just as in Bloch and Laneman~\cite{Bloch}, we consider six measures of asymptotic independence of varying strengths. We show that for many   classes of degraded, memoryless wiretap channels, the $\eps$-secrecy capacity (maximum code rate $R$ such that  the error probability is asymptotically no larger than $\eps$) does not depend on $\eps$. In other words, the $\eps$-secrecy capacity  is not larger than the expression in \eqref{eqn:sc} in which it is assumed that the error probability of decoding $M$ vanishes asymptotically.  Because we still ask that the leakage rate vanishes with the blocklength, we say that a {\em partial strong converse} holds. 

\subsection{Related Work}

\begin{figure}[t]
  \centering
   \begin{overpic}[width=3.5in]{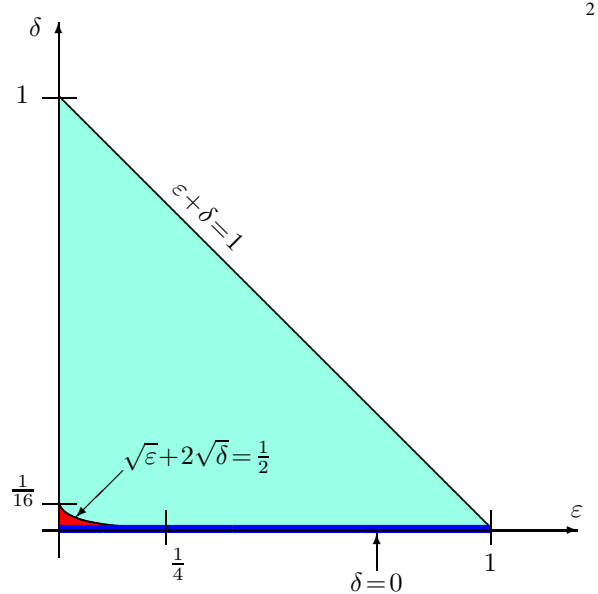}
     \put(19.4,5){\vector(0,1){80}}    
    \put(17,9){\vector(1,0){80}}   
    \put(96,11){$\eps$}    
    \put(15,83){$\delta$}    
    \put(36,60){\rotatebox{-45}{$\eps\! +\! \delta\! =\! 1$}}
     \put(29,19){{$\sqrt{\eps}\!+\!2\sqrt{\delta}\!=\!\frac{1}{2}$}}
         \put(29,18){\vector(-1,-1){7}}
    \put(63,0){$\delta\! =\! 0$} 
     \put(67,3){\vector(0,1){5.5}}
    \put(83,3){$1$} 
    \put(36,3){$\frac{1}{4}$} 
     \put(35.4,7){\line(0,1){5}}  
     \put(84.1,7){\line(0,1){5}}  
    \put(12,14){$\frac{1}{16}$} 
    \put(17,13){\line(1,0){5}}  
\put(17,73.8){\line(1,0){5}}
     \put(13,73){$1$} 
   \end{overpic}
     \caption{The strong converses for the degraded DM-WTC (Theorem \ref{thm:str_conv_dwtc}) and G-WTC  (Theorem \ref{thm:str_conv_gauss}) hold  for $(\eps,\delta)$ on  the  blue strip $[0,1)\times \{0\}$ (i.e., partial strong converse). Here, $\eps$ denotes the   error probability and $\delta$   the variational distance defined in \eqref{eqn:var_dist}. Morgan and Winter's \cite[Thm.~14]{morgan} pretty strong converse for the private capacity of degradable quantum  channels  holds for $\sqrt{\eps}+2\sqrt{ \delta}<1/2$, indicated by the region in red.  Tyagi and Watanabe's strong converse for the secret key capacity \cite[Cor.~11]{TW14}  holds for $\eps+\delta<1$, indicated by the union of the cyan, red and blue regions. This result improves  on Tyagi and Narayan's strong converse for the same problem~\cite[Sec.~VII]{TN13} which  holds for $(\eps,\delta)\in [0,1)\times \{0\}$.   We caution that these information-theoretic security problems are   different  so the results  are   not directly comparable.} 
     \label{fig:reg}
\end{figure}

In the   majority of the information-theoretic security literature~\cite{BlochBarros,Liang},   only {\em weak} converse statements are established, typically using Fano's inequality.  However,  some progress has been made in recent works to establish {\em strong} converses. For example,  the authors of~\cite{TN13,HTW14, TW14} proved strong converses for the multi-party secret key agreement problem and other related problems. In particular, the authors of \cite{HTW14, TW14} proved that the secret key capacity does not depend on the error bound $\eps$ and the secrecy bound $\delta$ (measured according to the variational distance) as long as $\eps+\delta<1$.  Another related work is the one by Morgan and Winter~\cite[Sec.~VI]{morgan} who used one-shot bounds in \cite{renes} to establish   a so-called {\em pretty strong converse} for the private capacity of a degradable quantum channel. Specifically, they prove that the that private capacity does not depend on the error bound $\eps$ and the secrecy bound  $\delta$  as long as  $\sqrt{\eps}+2\sqrt{ \delta}<1/2$.\footnote{\label{fn:var}To be more precise,  the authors in \cite[Thm.~14]{morgan}  used results in~\cite{renes} to  prove  that the private capacity does not depend on   $\eps'$ and $\delta'$, both measured in terms of the {\em purified distance} $d_{\mathrm{pur}}(P,Q)  := \big[ 1- (\sum_x ( P(x)Q(x)  )^{1/2} )^2\big]^{1/2}$, as long as $\eps'+2\delta'<1/\sqrt{2}$. This can be translated to the true average error probability $\eps$ and the  variational distance $\delta$ using  the bounds $\eps' \le \sqrt{2\eps}$ and $\delta'\le\sqrt{2\delta}$ (e.g.~\cite[Thm.~1]{fuchs}). Thus, one obtains the  strong converse  condition $\sqrt{\eps}+2\sqrt{ \delta}<1/2$  (albeit conservative)  in terms of the error probability and variational distance.  }   In the present paper, we only prove a strong converse for $(\eps,\delta) \in [0,1)\times\{0\}$ and a comparison of the results from various related works is shown in Fig.~\ref{fig:reg}. There is substantial motivation to prove  strong converses because  such   statements  indicate  that there exists a {\em sharp phase transition} between rates  that are achievable and those that are not.  Codes with unachievable rates have error probabilities that tend to one (or a positive number strictly less than one for the pretty strong converse) as the blocklength grows. Unlike weak converses, the rates  are not simply bounded away from zero. For point-to-point channel coding, Wolfowitz established the strong converse in the    1950s~\cite{Wolfowitz1957}, but little attention has been paid to strong converses for information-theoretic problems with secrecy constraints, such as the wiretap channel.

\newcommand\independent{\protect\mathpalette{\protect\independenT}{\perp}}
\def\independenT#1#2{\mathrel{\rlap{$#1#2$}\mkern2mu{#1#2}}}

\begin{figure}[t]
\centering
\setlength{\unitlength}{.4mm}
\begin{picture}(210, 90)
\put(0, 45){\vector(1, 0){30}}
\put(60, 45){\vector(1,0){30}}
\put(30, 30){\line(1, 0){30}}
\put(30, 30){\line(0,1){30}}
\put(60, 30){\line(0,1){30}}
\put(30, 60){\line(1,0){30}}


\put(90, 0){\line(1, 0){30}}
\put(90, 0){\line(0,1){90}}
\put(120, 0){\line(0,1){90}}
\put(90, 90){\line(1,0){30}}

\put(0, 50){  $M$}
\put(66, 50){  $X^n$}
\put(32, 42){$Q_{X^n|M}$ } 
\put(99, 42){$W^n$} 

\put(126, 20){  $Z^n$ almost $\independent$ of $M$}
\put(126, 80){  $Y^n$}

\put(186, 80){  $\hatM$}

\put(120, 15){\vector(1, 0){30}}
\put(120, 75){\vector(1, 0){30}}



\put(150, 60){\line(1, 0){30}}
\put(150, 60){\line(0,1){30}}
\put(180, 60){\line(0,1){30}}
\put(150, 90){\line(1,0){30}}

\put(180, 75){\vector(1, 0){30}}

\put(161, 72){$\varphi_n$ } 


  \end{picture}
  \caption{Illustration of the wiretap channel model. The decoding error probability must satisfy $ \limsup_{n\to\infty}\Pr(\hatM\ne M)\le\eps$ while the leakage $\bbS_i(P_{MZ^n}, P_M\times P_{Z^n})$ (measured according to any one of the six secrecy metrics in Definition~\ref{def:metrics}) must vanish as $n$ grows.   }
  \label{fig:model}
\end{figure}
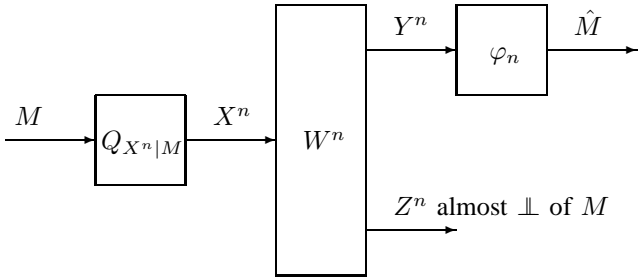

\subsection{Summary Of Our Approach}
In this work, we adopt the information spectrum method~\cite{VH94, HV93, Han10} to make strong converse statements for various classes of degraded wiretap channels. The information spectrum method, developed by Verd\'u and Han~\cite{VH94, HV93}, is a systematic and powerful method to characterize the fundamental limits of communication  systems without  the usual assumptions of memorylessness, stationarity, ergodicity and information stability. The information spectrum method is also useful in  establishing necessary and sufficient conditions for the strong converse to hold~\cite[Sec.~3.5]{Han10} which  is one of the reasons why  we have adopted this approach.  The use of the information spectrum approach for general  wiretap channels was pioneered by Hayashi in~\cite{Hayashi06} in which the use of channel resolvability~\cite{HV93}~\cite[Sec.~6.3]{Han10} was shown to be a  useful coding mechanism for secrecy.

\subsection{Main Contributions}
In this work, we establish that the degraded discrete memoryless (and stationary) wiretap channel  (DM-WTC) admits a partial  strong converse. This means that regardless of the permissible asymptotic  error probability $\eps \in [0,1)$, if the leakage vanishes asymptotically, the maximum rate of transmission cannot exceed the secrecy capacity Wyner   derived in~\eqref{eqn:sc}. This contribution is a strengthening of Wyner's seminal result~\cite{Wyn75}. We extend our result to prove the same for the    Gaussian wiretap channel (G-WTC), strengthening the capacity result by Leung-Yan-Cheong and Hellman  \cite{leung}. Finally, we prove that the partial strong converse holds for some classes of non-stationary wiretap channels.

\subsection{Paper Organization}
This paper is organized as follows. In Section~\ref{sec:defs}, we state the notational conventions, describe the system model and     define the partial strong converse property for the wiretap channel. In Section~\ref{sec:main_res}, we state our main results. In particular, after recapitulating some information spectrum quantities in Section \ref{sec:info_spec}, we state   general formulas for the $\eps$-secrecy capacity and its optimistic version in Section~\ref{sec:gen_form}. These are done for  arbitrary wiretap channels where the legitimate receiver is allowed to make an error with probability not exceeding $\eps \in [0,1)$ but the leakage is required to tend to zero. The bulk of the contributions is contained in Section~\ref{sec:str_conve} where we present strong converse results for specific channel models such as the DM-WTC and the G-WTC. We conclude and suggest avenues for future research in Section \ref{sec:concl}.  The proofs of the theorems are contained in Section~\ref{sec:prfs}.

\section{System Model and Definitions} \label{sec:defs}
In this section, we state our notation and the definitions of the various problems we consider in this paper. 
\subsection{Basic Notations}

Random variables (e.g., $X$) and their realizations (e.g., $x$) are denoted by upper case and lower case serif font, respectively. 
Sets are denoted in calligraphic font (e.g., the alphabet of $X$ is $\calX$). We use the notation $X^n$ to denote a vector of random 
variables $(X_1, \ldots , X_n)$. In addition, $\bX = \{ X^n \}_{n\in\bbN}$  is a {\em general source} in the sense that each member of the sequence $X^n = (X^{(n)}_1, \ldots   , X^{(n)}_n)$ is a random vector. The consistency condition (i.e., $X^{(n)}_i = X^{(m )}_i$ for all $m < n$ and  $1\le i\le m$) need not hold.  A {\em general broadcast channel} $\bW = \{W^n : \calX^n\to\calY^n\times\calZ^n\}_{n\in\bbN}$   is a sequence of stochastic mappings from $\calX^n$ to $\calY^n\times\calZ^n$. The  set of all probability distributions with support on an alphabet $\calX$ is denoted as $\calP(\calX)$. We use the notation $X\sim P_X$ to mean  that the distribution of $X$ is $P_X$. The joint  distribution formed by the product of the input distribution  $P_X \in\calP(\calX)$ and the channel $W:\calX\to\calY$ is denoted by $P_X\times W$. Information-theoretic quantities are denoted using   the   notations in Han's book~\cite{Han10}, e.g., $H(X)$ for entropy, $I(X ; Y )$ for mutual information and $D(P\| Q)$ for relative entropy. All logarithms are to an arbitrary base. We also   use the discrete interval  notation $[i : j] := \{i, \ldots , j\}$. 

The {\em variational distance} between two measures or distributions $P$ and $Q$ on the same space $\calX$ is defined as
\begin{equation}
\bbV(P,Q) :=  \sup_{\calA\subset\calX} \big| P(\calA)-Q(\calA)\big| ,\label{eqn:var_dist}
\end{equation}
where $\calA\subset\calX$ runs over the class of measurable subsets of $\calX$. 
For an arbitrary sample space $\calX$, the definition of the variational distance in \eqref{eqn:var_dist} is equivalent to
\begin{equation}
\bbV(P,Q) = \frac{1}{2} \int_\calX  \big| p(x)-q(x) \big| \, \rmd\lambda(x),\label{eqn:var_dist2}
\end{equation}
where $\lambda$ is a common dominating measure of $P$ and $Q$ and $p(x) = \rmd P/\rmd\lambda$ and $q(x) = \rmd Q/\rmd\lambda$ denote their respective densities.  Furthermore, $\bbV(P,Q) = P(\calA^*)-Q(\calA^*)$ where $\calA^*=\{x: p(x)\ge q(x)\}$.  

The probability density function of the normal distribution $\calN(y;\mu,\sigma^2)$ is defined as 
\begin{equation}
\calN(y;\mu,\sigma^2):=\frac{1}{\sqrt{2\pi\sigma^2} }\exp\bigg( -\frac{(y-\mu)^2}{2\sigma^2}\bigg).
\end{equation}
\subsection{System Model}
We consider a general wiretap channel, which is simply a general broadcast channel   $\bW= \{W^n : \calX^n\to\calY^n\times\calZ^n\}_{n\in\bbN}$. Terminal $\calX$ denotes the sender, terminal $\calY$ denotes the legitimate receiver, and terminal $\calZ$ denotes the eavesdropper. We would like to reliably transmit a message $M$ from the terminal $\calX$ to terminal $\calY$, and at the same time, design the code such that terminal $\calZ$, the eavesdropper, obtains no information about $M$. More precisely, the eavesdropper's signal or observation $Z^n$ is required to be {\em asymptotically independent} of $M$. There are various ways to quantify asymptotic independence. We adopt the methodology of Bloch and Laneman~\cite{Bloch} and consider six   metrics of varying strengths that quantify asymptotic independence.
\begin{definition} \label{def:metrics}
Let $\eta>0$ be an arbitrary constant, $P:=P_{MZ^n}$, $Q:=P_M \times P_{Z^n}$ and $(M,Z^n)\sim P$.  Consider the following measures of independence, also known as {\em secrecy metrics}:
\begin{align}
\bbS_1(P,Q)  & :=  D( P  \| Q), \\
\bbS_2(P,Q)  & :=  \bbV( P  , Q),   \label{eqn:var_leak} \\ 
\bbS_3^\eta(P,Q)  & :=  \Pr \Big(  \log\frac{P  (M, Z^n)}{Q(M,Z^n)}  > \eta \Big), \\
\bbS_4(P,Q)  & :=   \frac{1}{n} D( P  \| Q) , \\
\bbS_5(P,Q)  & := \frac{1}{n}\bbV( P  , Q) ,   \\ 
\bbS_6^\eta (P,Q) & :=  \Pr \Big( \frac{1}{n} \log\frac{P  (M, Z^n)}{ Q(M,Z^n)}  > \eta\Big)  .  \label{eqn:S6}
\end{align} 
\end{definition}
Because $D(P\|Q)=D( P_{MZ^n} \| P_{M}  \times   P_{Z^n})=I(M;Z^n)$ the mutual information,  secrecy metrics $\bbS_1$ and $\bbS_4$ correspond to strong \cite{Maurer:2000, csi96} and weak secrecy~\cite{Wyn75} respectively. These are the most common metrics in the information-theoretic security literature~\cite{BlochBarros, Liang}. 
We say that $\bbS_i$  {\em dominates} $\bbS_j$  if $\bbS_i( P_{MZ^n}, P_{M} \times P_{Z^n})\to 0$ implies that $\bbS_j( P_{MZ^n}, P_{M} \times P_{Z^n}) \to 0$ and we denote this  by $\bbS_i\succeq\bbS_j$. Bloch and Laneman~\cite[Prop.~1]{Bloch}  showed that there exists an ordering of the above six secrecy metrics. In particular, for any $\eta_1,\eta_2>0$, 
\begin{equation}
\bbS_1 \succeq \bbS_2 \succeq  \bbS_3^{\eta_1} \succeq \bbS_4 \succeq \bbS_5 \succeq \bbS_6^{\eta_2}  . \label{eqn:order}
\end{equation}

Given the wiretap channel $\bW= \{W^n\}_{n\in\bbN}$, we define its $\calY^n$- and {\em $\calZ^n$-marginals} as 
\begin{align}
W^n_{\calY}(\by|\bx) & := \sum_{\bz \in\calZ^n} W^n(\by,\bz|\bx),  \quad \mbox{and}\\
 W^n_{\calZ}(\bz|\bx) &:= \sum_{\by \in\calY^n} W^n(\by,\bz|\bx),
\end{align}
where $(\bx,\by,\bz)\in\calX^n\times\calY^n\times\calZ^n$ is a tuple of vectors.

\begin{definition}
An {\em $(n,M_n,\eps_n ,\delta_n)$-wiretap code for secrecy metric $ i \in [1:6]$} consists of  (see Fig.~\ref{fig:model})
\begin{enumerate}
\item A message set $\calM_n= [ 1:M_n]$;
\item A stochastic encoder $Q_{X^n|M } : \calM_n \to\calX^n$  and
\item A decoder $\varphi_n:\calY^n\to \calM_n$
\end{enumerate}
such that the average error probability satisfies
\begin{equation}
\frac{1}{M_n}\sum_{m\in\calM_n} \sum_{\bx\in\calX^n} Q_{X^n|M  }(\bx|m)  W_{\calY}^n(\calY^n\setminus\varphi_n^{-1}(m) |\bx) \le \eps_n \label{eqn:decoding_err}
\end{equation}
and the information leakage   satisfies
\begin{equation}
\bbS_i( P_{MZ^n}, P_{M} \times P_{Z^n}) \le   \delta_n \label{eqn:leakage_rate}
\end{equation}
where $M \in\calM_n$ is the message random variable which is uniformly distributed over  $\calM_n$. 
\end{definition}

We remark that secrecy metrics $\bbS_3^\eta$ and $\bbS_6^\eta$ depend on an additional parameter $\eta>0$ but to simplify notation, we do not make the dependence of the code on $\eta$ explicit. This should not cause any confusion in the sequel.

We now define achievable rates and capacities for the general wiretap channel. 

\begin{definition} \label{def:cap}
Let  $\eps\in [0,1)$ and $i \in [1:6]$. Let $R \in\bbR$ be called an {\em $(\eps,i)$-achievable rate} for the general wiretap channel $\bW$ if there exists a sequence of $(n,M_n,\eps_n ,\delta_n)$-wiretap codes   for secrecy metric $i$  such that 
\begin{equation}
\limsup_{n\to\infty}\,  \eps_n\le\eps,\,\,  \lim_{n\to\infty}\delta_n=0,\,\,\mbox{and}\,\, \liminf_{n\to\infty}\,  \frac{1}{n}\log M_n\ge R. \label{eqn:pess_def}
\end{equation}
Define the {\em  $(\eps,i)$-secrecy capacity} (or simply {\em $(\eps,i)$-capacity}) of the wiretap channel $\bW$ as 
\begin{equation}
C_\eps^{(i)}(\bW) := \sup\{R : R \mbox{ is }(\eps,i)\mbox{-achievable}\}.
\end{equation} 
Define the {\em $(i)$-secrecy capacity} (or simply {\em $(i)$-capacity})  of the wiretap channel   $\bW$ as 
\begin{equation}
 C^{(i)}(\bW) :=C_0^{(i)}(\bW). \label{eqn:pess_cap}
\end{equation} 
\end{definition}
We note that the error probability  is allowed to be any number in $[0,1)$ but the secrecy metric is required to tend to zero as the blocklength grows. From the ordering of the secrecy metrics in \eqref{eqn:order}, we know that for every $\eps \in [0,1)$, we have 
\begin{equation}
C^{(i) }_\eps (\bW)  \le C^{(j)}_\eps (\bW),\quad\mbox{if}\quad i\le j. \label{eqn:order_pess}
\end{equation}

For the definition of the partial strong converse property, we find it useful to first consider {\em optimistic} analogues \cite[Def.~3.9.1]{Han10} \cite[Thm.~4.3]{chen99} \cite[Thm.~7]{Ste98} of fundamental limits, such as the capacity in Definition \ref{def:cap}. 

\begin{definition} \label{def:opt}
Let  $\eps\in (0,1]$ and $i \in [1:6]$. Let $R\in\bbR$ be called an {\em $(\eps,i)$-optimistically  achievable  rate} for the general wiretap channel $\bW$ if  for all sequences of $(n,M_n,\eps_n ,\delta_n)$-wiretap codes for secrecy metric $i$  satisfying
\begin{equation}
\liminf_{n\to\infty}\,  \frac{1}{n}\log M_n\ge R\quad\mbox{and}\quad \lim_{n\to\infty}\delta_n=0,\label{eqn:opt_def1}
\end{equation}
we must  also have 
\begin{equation}
\liminf_{n\to\infty}\,  \eps_n \ge \eps.
\end{equation}
Define the {\em  $(\eps,i)$-optimistic secrecy capacity} (or simply {\em $(\eps,i)$-optimistic capacity})  of the wiretap channel  $\bW$  as 
\begin{equation}
\overC_\eps^{(i)}(\bW) := \inf\{R : R \mbox{ is }(\eps,i)\mbox{-optimistically  achievable}\}.
\end{equation}
Define the {\em $(i)$-optimistic secrecy capacity}  (or simply {\em $(i)$-optimistic capacity})   of the wiretap channel   $\bW$ as 
\begin{equation}
 \overC^{(i)}(\bW) :=\overC_1^{(i)}(\bW) . \label{eqn:opt_cap}
\end{equation} 
\end{definition}

Following \cite{TomTan13a}, and by contrapositive, we note that the $(\eps,i)$-optimistic capacity can equivalently be defined as the supremum of all numbers $R\in\bbR$ for which there exists a sequence  of $(n,M_n,\eps_n ,\delta_n)$-wiretap codes   for secrecy metric $i$ such that 
\begin{equation}
\liminf_{n\to\infty}\,  \eps_n < \eps,\,\, \lim_{n\to\infty}\delta_n=0,\,\,\mbox{and}\,\, \liminf_{n\to\infty}\,  \frac{1}{n}\log M_n\ge R. \label{eqn:opt_def2}
\end{equation}
The first condition in \eqref{eqn:opt_def2} explains the term {\em optimistic}. Indeed, by the definition of $\liminf$ the error probability is only required to be smaller than $\eps$ {\em for  infinitely many $n$} as opposed to  {\em for all sufficiently large $n$}, implied by the first condition in \eqref{eqn:pess_def} for the (pessimistic) capacity.  Note that our definition of the optimistic capacity  in Definition~\ref{def:opt}, or equivalently the conditions in \eqref{eqn:opt_def2}, is slightly different from  those in Chen and Alajaji~\cite[Def.~4.9]{chen99} and Steinberg~\cite[Thm.~7]{Ste98}. 
Our definition has the advantage that it  allows us to characterize the $(\eps,i)$-optimistic secrecy capacity as an  {\em equality} for all $\eps \in (0, 1]$. We refer the reader to \cite[Sec.~IV]{VH94} and \cite[Rmk.~1.6.3]{Han10} for a discussion of this subtlety.   

From the ordering of the secrecy metrics in \eqref{eqn:order}, we know that for every $\eps \in (0,1]$, we have 
\begin{equation}
\overC^{(i) }_\eps (\bW)  \le \overC^{(j)}_\eps (\bW),\quad\mbox{if}\quad i\le j.  \label{eqn:order_over}
\end{equation}
 It is also easily seen from the definitions that  for all $i \in [1:6]$,
\begin{equation}
 C^{(i)}(\bW)\le \overC^{(i)}(\bW).\label{eqn:over}
 \end{equation} 
Equality in \eqref{eqn:over} is particularly significant as can be seen from the following definition.

\begin{definition} \label{def:str_con}
A wiretap channel  $\bW$  is said to satisfy the {\em partial strong converse under secrecy metric $i \in [1:6]$}  if 
\begin{equation}
C^{(i)}(\bW) =\overC^{(i)}(\bW) . 
\end{equation}
\end{definition}
The qualifier {\em partial} is used because we still insist that the information leakage, represented by $\delta_n$, tends to zero. The strong converse thus only pertains to the probability of decoding error in \eqref{eqn:decoding_err}.  This definition of the partial strong converse corresponds to that presented by Han~\cite[Sec.~3.7]{Han10} and Hayashi and Nagaoka in \cite{Hayashi03}. Clearly, if $\bW$ satisfies the partial strong converse under secrecy metric $i$, both $C_\eps^{(i)}(\bW)$ and $\overC_\eps^{(i)}(\bW)$ do not depend on $\eps$. More precisely, the partial strong converse implies that
\begin{align}
C_\eps^{(i)}(\bW) &=C^{(i)}(\bW) \quad\forall\, \eps\in[0,1) \quad\mbox{and} \label{eqn:weak1}\\
\overC_\eps^{(i)}(\bW) &= \overC^{(i)}(\bW) \quad\forall\, \eps\in(0,1]\label{eqn:weak2} .
\end{align}
 However, as discussed in \cite[Rmk.~3.5.1]{Han10}, Definition~\ref{def:str_con} implies~\eqref{eqn:weak1}--\eqref{eqn:weak2} but not the other way round. 

\section{Main Results} \label{sec:main_res}
In this section, we state our results. First, we generalize the results in \cite{Bloch, Hayashi06} and characterize $C_\eps^{(i)}(\bW)$  and $\overC_\eps^{(i)}(\bW) $ for general wiretap channels. We then state our main result, namely that degraded DM-WTC  admit the partial strong converse. We also show that certain classes of non-stationary wiretap channels and the Gaussian wiretap channel possess  the partial strong converse property.

\subsection{Basic Quantities in   Information Spectrum Analysis}\label{sec:info_spec}
To state our results concisely, we recall some definitions from information spectrum analysis \cite{Han10,chenAlajaji}.  For a general sequence of random variables $\bB = \{B_n\}_{n \in\bbN}$, define 
\begin{align}
\eps\text{-}\pliminf_{n\to\infty} \!  B_n  \!:=\! \sup \Big\{ r : \limsup_{n\to\infty}  \Pr ( B_n\le r )\! \le\! \eps \Big\}  \label{eqn:epliminf}
\end{align}
for $\eps \in [0,1)$,  and 
\begin{align}
\eps\text{-}\plimsup_{n\to\infty} \!  B_n  \!:=\! \sup \Big\{r : \liminf_{n\to\infty}  \Pr ( B_n \le r ) \!<\! \eps \Big\}  \label{eqn:eplimsup}
\end{align}
for $\eps \in (0,1]$.
Notice the strict inequality in  \eqref{eqn:eplimsup}, which differs from the non-strict inequality in~\eqref{eqn:epliminf}. The $\pliminf$ and $\plimsup$ are defined as $0\text{-}\pliminf$ and $1\text{-}\plimsup$ respectively. 
 For any general   pair of random variables $(\bV,\bY)$ with joint distribution  $P_{\bV \bY} := \{ P_{V^n  Y^n} \}_{n\in\bbN}$, define, for each $n$, the normalized information density random variables  
\begin{equation}
\imath_{n} (V^n;Y^n) := \frac{1}{n}\log\frac{P_{Y^n|V^n}(Y^n|V^n)}{P_{Y^n}(Y^n)}.
\end{equation}
Given  $\{\imath_{n} (V^n;Y^n)\}_{n\in\bbN}$, we may now define
\begin{align}
 \underI_\eps (\bV;\bY) &:=\eps\text{-}\pliminf_{n\to\infty}\, \imath_{ n} (V^n;Y^n)\label{eqn:underIeps} \\
  \overI_\eps (\bV;\bY) &:=\eps\text{-}\plimsup_{n\to\infty} \imath_{ n} (V^n;Y^n) \label{eqn:overIeps} 
\end{align}
The properties of $\underI_\eps (\bV;\bY)$ and $\overI_\eps (\bV;\bY)$ are described in \cite[Sec.~2.4]{chenAlajaji}.
When $\eps=0$ in \eqref{eqn:underIeps} and $\eps=1$ in \eqref{eqn:overIeps}, we leave out the subscripts, i.e.,  we  define 
\begin{equation}
 \underI (\bV;\bY) := \underI_0(\bV;\bY),\quad\mbox{and} \quad \overI(\bV;\bY) :=\overI_1 (\bV;\bY).
\end{equation}
In information spectrum analysis, $\underI (\bV;\bY)$ and $\overI (\bV;\bY)$
 are termed the {\em spectral inf- and sup-mutual information rates} respectively. They are respectively the $\pliminf$ and $\plimsup$ of  the sequence of  random variables $\{\imath_{ n} (V^n;Y^n)\}_{n\in\bbN}$.

\subsection{Capacity and Strong Converse Results for General Wiretap Channels} \label{sec:gen_form}

The following theorem is a straightforward extension  of the results in \cite{Bloch, Hayashi06}. For completeness, a proof sketch is provided in Section~\ref{sec:prf_cap}.
\begin{theorem}[General Formula] \label{prop:cap} 
For $i \in [2:6]$, the $(\eps, i)$-capacity and  the  $(\eps, i)$-optimistic capacity of any general wiretap channel $\bW$ are
\begin{align}
C_\eps^{(i)} (\bW)  &= \sup_{\bV - \bX-(\bY,\bZ)}  \underI_\eps (\bV;\bY) - \overI(\bV;\bZ), \quad\mbox{and}  \label{eqn:Ceps} \\
\overC_\eps^{(i)} (\bW)  &=  \sup_{\bV-\bX-(\bY,\bZ)}  \overI_\eps (\bV;\bY) - \overI(\bV;\bZ).\label{eqn:Ceps_dagger} 
\end{align}
The suprema are over the set of all sequences of   distributions $P_{\bV \bX } = \{P_{V^n X^n }\}_{n\in\bbN}$ or equivalently over all Markov chains\footnote{The notation $\bA-\bB-\bC$ means that $A^n-B^n-C^n$ forms a Markov chain for all $n\in\bbN$. } $\bV-\bX-(\bY,\bZ)$ where the distribution of $(\bY,\bZ)$ given $\bX$ corresponds to the wiretap channel $\bW$. 
\end{theorem}
Using the definition of the partial strong converse in Definition \ref{def:str_con}, we immediately obtain the following corollary, applicable {\em only} to secrecy metrics $\bbS_i,i\in [2:6]$.
\begin{corollary}[General Partial Strong Converse]
For any wiretap channel $\bW$ and any secrecy metric $\bbS_i,i\in [2:6]$, the partial strong converse property  holds   if and only if 
\begin{equation}
 \sup \underI  (\bV;\bY) - \overI(\bV;\bZ)=  \sup  \overI  (\bV;\bY) - \overI(\bV;\bZ), \label{eqn:str_conv}
\end{equation}
where the suprema are understood to be the same as in Theorem \ref{prop:cap}.
\end{corollary}

\subsection{Strong Converse Theorems for Specific Wiretap Channel Models}\label{sec:str_conve}
\subsubsection{Degraded Discrete Memoryless Wiretap Channels}
A   {\em physically degraded},   or simply {\em  degraded}, wiretap channel $\bW$ is one in which  for every $n\in\bbN$, and for every $(\bx,\by,\bz)\in\calX^n\times\calY^n\times\calZ^n$,
\begin{equation}
W^n(\by,\bz|\bx)=W^n_1(\by |\bx) W^n_{2}(\bz |\by) \label{eqn:degraded}
\end{equation}
 for some channels $W_1^n:\calX^n\to\calY^n$ and $W_{2}^n:\calY^n\to\calZ^n$.  In other words, $X^n-Y^n-Z^n$ forms a Markov chain for every $n \in\bbN$.  A DM-WTC has alphabets $\calX,\calY,\calZ$ that are finite sets and the channel is  {\em stationary} and {\em memoryless}  in the sense that 
\begin{equation}
W^n_1(\by |\bx)=\prod_{i=1}^n W_1(y_i|x_i),\,\,\mbox{and}\,\,  W^n_{2}(\bz |\by)=\prod_{i=1}^n W_{ 2}(z_i|y_i) \label{eqn:memoryless}
\end{equation}
for every $(\bx,\by,\bz)\in\calX^n\times\calY^n\times\calZ^n$.  It is  known from Wyner's seminal work on the wiretap channel~\cite{Wyn75} that the capacity of a degraded DM-WTC $W:\calX\to\calY\times\calZ$ (under the weak secrecy criterion $\bbS_4$) is 
\begin{equation}
\CDM(W) := \max_{P_X } I(X;Y|Z) = \max_{P_X} I(X;Y)-I(X;Z) , \label{eqn:cap_dmtc}
\end{equation}
where the mutual information quantities are calculated according to $\Pr(Y=y,Z=z| X=x )=W(y,z|x)$.  The second equality in \eqref{eqn:cap_dmtc} follows from the fact that $X-Y-Z$ forms a Markov chain (degradedness) so $I(X;Y|Z )=I(X;YZ)-I(X;Z)= I(X;Y )-I(X;Z)$.  
Wyner's weak converse~\cite[Eq.~(35) in Sec.~IV]{Wyn75} assumes that the probability of decoding error vanishes asymptotically.   The first of our main results is a strengthening of Wyner's seminal result.
\begin{theorem}[Degraded DM-WTCs] \label{thm:str_conv_dwtc}
Any degraded DM-WTC $W:\calX\to\calY\times\calZ$  satisfies the partial strong converse under any secrecy metric  $\bbS_i,i \in [1:6]$. Consequently, the $(i)$-capacities and $(i)$-optimistic capacities of $\bW=\{W\}$ are equal to $\CDM(W)$ for all $i\in [1:6]$.
\end{theorem}
A proof of this theorem is provided in  Section~\ref{sec:prf_degraded}. The basic idea is to lower bound the $(1)$-capacity $C^{(1)}(\bW)$ (capacity under secrecy metric $\bbS_1$) with $\CDM(W)$ and to upper bound the  $(6)$-optimistic capacity $\overC^{(6)}(\bW)$  (optimistic capacity under secrecy metric $\bbS_6$) with the same quantity $\CDM(W)$. This then allows us to assert  that $C^{(1)}(\bW) = \overC^{(6)}(\bW)$ showing from \eqref{eqn:order_pess}, \eqref{eqn:order_over} and \eqref{eqn:over} that $C^{(i)}(\bW) = \overC^{(i)}(\bW)$ for all $i\in [1:6]$, i.e., the partial strong converse holds under all $6$ secrecy metrics. The lower bound of  $C^{(1)}(\bW)$ is straightforward and follows by using the connection between secrecy and channel resolvability \cite{Hayashi06,Bloch}, independent and identically distributed  (i.i.d.) random codes and standard (large deviation) concentration bounds~\cite{Dembo}. This sequence of steps is already well known. See for example~\cite[Remark~3]{Bloch} or the papers by Hayashi \cite{Hayashi11, Hayashi06} and Han {\em et al.} \cite{Han2013} on secrecy and reliability exponents for the wiretap channel.  The interesting part of the proof  is in the upper bound of  
\begin{equation}
\overC^{(6)}(\bW) = \sup_{\bV-\bX-(\bY,\bZ)}\overI(\bV;\bY)- \overI(\bV;\bZ). \label{eqn:C6}
\end{equation}
The difficulty arises because  we need to upper bound and subsequently single-letterize the supremum of the difference between two limit superiors in probability. To perform these tasks, we leverage the proof technique for \cite[Thm.~3.7.2]{Han10} and combine several  known results and techniques from the information-theoretic security literature.


\subsubsection{Non-Stationary Wiretap Channels}
The assumption of degradedness in Theorem~\ref{thm:str_conv_dwtc} is rather strong but appears essential in the proof. We do not think that the assumption concerning memorylessness is critical (cf.~\cite[Cor.~3]{TomTan13a}),  but we defer the study of wiretap channels with memory to future work. Instead we examine   conditions under which the stationarity assumption may be relaxed. In this section, we assume that the wiretap channel is degraded in the sense of~\eqref{eqn:degraded} but  the components have the following non-stationary structure:
\begin{equation}
W^n_1(\by |\bx)=\prod_{i=1}^n W_{1i}(y_i|x_i),\,\,\mbox{and}\,\,  W^n_{2}(\bz |\by)=\prod_{i=1}^n W_{ 2i}(z_i|y_i). \label{eqn:non_stat}
\end{equation}
That is, the channels themselves may differ across time but the channel noises are nonetheless independent. We define the $i$-th wiretap channel as $W_i(y,z|x) := W_{1i}(y|x) W_{2i}(z|y)$. The main and eavesdropper's channels    are defined as $W_{\calY, i}(y|x):= W_{1i}(y|x)$ and $W_{\calZ,i} (z|x) :=\sum_{y\in\calY} W_{1i}(y|x)W_{2i}(z|y)$
respectively. 
These channels have Shannon capacities $C(W_{\calY,i})$ and $C(W_{\calZ,i})$ respectively.  We further assume that all component channels $\{W_{\calY,i}\}_{i\in\bbN}$ and $\{ W_{\calZ,i}\}_{i\in\bbN}$ are {\em weakly symmetric} \cite[Def.~3.4]{BlochBarros}. Recall that a discrete memoryless channel $V:\calX\to\calY$ is  weakly symmetric    if the rows of the channel transition probability matrix are permutations of each other and the column sums $\sum_{x\in\calX} V(y|x)$ are independent of $y$. Under the condition that the channels are degraded and weakly symmetric, Leung-Yan-Cheong \cite{lyc}  (also see \cite[Prop.~3.2]{BlochBarros}) showed that the secrecy capacity is the difference of the   capacities of the main and eavesdropper's channels, i.e., 
\begin{equation}
\CDM(W_i) = C(W_{\calY,i})-C(W_{\calZ,i}). \label{eqn:diff_cap}
\end{equation}
Note that \eqref{eqn:diff_cap} is a consequence of the fact that the (unique) capacity-achieving input distributions of the channels $W_{\calY,i}$ and $W_{\calZ,i}$  are the same and, in particular, they are uniform on $\calX$. See van Dijk~\cite{vanDijk} for further discussions. With these preparations, we are in a position to state the following result:

\begin{theorem}[Non-Stationary Wiretap Channels] \label{thm:str_conv_non}
Consider the degraded, discrete, memoryless but non-stationary wiretap channel in~\eqref{eqn:non_stat}. Assume that all $\{W_{\calY,i}\}_{i\in\bbN}$ and $\{ W_{\calZ,i}\}_{i\in\bbN}$ are    weakly symmetric  channels.  
  Under any secrecy metric  $\bbS_i,i\in [1:6]$,  the partial strong converse holds for $\bW=\{W^n\}_{n\in\bbN}$ if   the following limits exist: 
\begin{align}
  \lim_{n\to\infty}\frac{1}{n}\sum_{i=1}^n C(W_{\calY,i}),\quad  \mbox{and}\quad   \lim_{n\to\infty} \frac{1}{n}\sum_{i=1}^n C(W_{\calZ,i}) \label{eqn:cesaro_means} .
\end{align}  
\end{theorem}
The proof of this theorem can be found in Section~\ref{sec:prf_non_stat}.

For the purposes of comparison,   consider a point-to-point, discrete,  memoryless and  non-stationary channel $V^n(\by|\bx) = \prod_{i=1}^n V_i(y_i|x_i)$. Let $P_{\barY_i}$ be the unique  \cite[Cor.~2 to Thm.~4.5.2]{gallagerIT} capacity-achieving output distribution of $V_i$. It satisfies $P_{\barY_i}(y)>0$ for all $y\in\calY$ if all outputs are reachable~\cite[Cor.~1 to Thm.~4.5.2]{gallagerIT}. Further assume that\footnote{This condition is automatically satisfied by weakly symmetric channels as $P_{\barY_i}(Y)=1/ |\calY|$  with probability one for all $i$. Indeed, \eqref{eqn:sup_var} is satisfied if the minimum values of the capacity-achieving output distributions are uniformly bounded away from zero, i.e.,  $\inf_{i\in\bbN}\min_{y \in\calY} P_{\barY_i} (y)>0$. }
\begin{equation}
\sup_{i\in\bbN}\, \max_{x\in\calX}\,\,\var\left[ \log\frac{V_i(Y | x) }{P_{\barY_i} (Y)}\right]<\infty. \label{eqn:sup_var}
\end{equation}
Then, it is easy to show from the strong converse theorem for general channels~\cite[Thm.~3.5.1]{Han10} and the relation between limits in probability and usual limits~\cite[Thm.~3.5.2]{Han10} that the strong converse for $\bV:= \{V^n=\prod_{i=1}^n V_i\}_{n\in\bbN}$ holds if and only if 
\begin{equation}
\liminf_{n\to\infty}\,   \frac{1}{n}\sum_{i=1}^n C(V_{ i})=\limsup_{n\to\infty}\,   \frac{1}{n}\sum_{i=1}^n C(V_{ i}). \label{eqn:p2p}
\end{equation}
In other words,  $\lim_{n\to\infty}\frac{1}{n}\sum_{i=1}^n C(V_{ i})$ exists.  Indeed, the left-hand-side of \eqref{eqn:p2p} is the capacity~\cite[Rmk.~3.2.3]{Han10} of the channel $\bV$,  while if we assume~\eqref{eqn:sup_var}, the right-hand-side is the optimistic capacity (a statement generalizing \cite[Example~5.14]{chenAlajaji}).  Thus, the equivalent condition in terms of the existence of the limits of the Ces\`aro means   $\frac{1}{n}\sum_{i=1}^n C(W_{\calY,i})$  and  $\frac{1}{n}\sum_{i=1}^n C(W_{\calZ,i})$  in Theorem~\ref{thm:str_conv_non} is a      generalization  of channels without secrecy constraints to degraded (but weakly symmetric) wiretap channels.  

Of course, the existence of the two limits is only a {\em sufficient} condition for the partial strong converse to hold. It appears to be rather challenging to assert that it is also {\em necessary}, or to find an alternate (and stronger) characterization that is both necessary and sufficient. 

\subsubsection{Gaussian Wiretap Channels} \label{sec:gauss}
 We now demonstrate that the assumption of discreteness in Theorem \ref{thm:str_conv_dwtc} is not  critical. In fact, we can make a partial strong converse statement for the   (memoryless, stationary) G-WTC  in which  all the alphabets are the real line $\bbR$ and the channel laws are 
\begin{align}
W_\calY(y|x)   := \calN(y;x,\sigma_1^2) ,\,\, \mbox{and}\,\,
W_\calZ(z|x)   := \calN(z;x,\sigma_2^2) ,
\end{align}
and we assume that $\sigma_2>\sigma_1$. Observe that by defining $W_1(y|x)=W_\calY(y|x)$ and 
$
W_{2}(z|y):=\calN(z;y,\sigma^2_2-\sigma_1^2)$, 
we see that the G-WTC  $W(y,z|x) = W_1(y|x)W_{2}(z|y)$ is degraded. In fact, to be more precise, it is {\em stochastically degraded} but we will not differentiate between physical degradedness and stochastic degradedness since the capacities and optimistic capacities are identical, a direct consequence of~\cite[Lem.~2.1]{Liang}. For every blocklength $n\in\bbN$, the input codeword $X^n$ is required to satisfy the almost sure power constraint
\begin{equation}
\Pr(X^n \in \calF_n) = 1 \label{eqn:cost}
\end{equation}
where 
\begin{equation}
\calF_n := \{\bx\in\bbR^n: \|\bx\|_2^2\le nS\}
\end{equation}
is the $(n-1)$-sphere with radius $\sqrt{nS}$ and $S>0$ is the  permissible power.
Recall from Leung-Yan-Cheong and Hellman  \cite{leung} that the capacity of the G-WTC, under secrecy metric $\bbS_4$  (weak secrecy) and assuming that the probability of decoding error vanishes asymptotically, is
\begin{equation} 
\CG(W;S) := \frac{1}{2}\log\bigg( 1+\frac{S}{\sigma_1^2}\bigg)- \frac{1}{2}\log\bigg( 1+\frac{S}{\sigma_2^2}\bigg).
\end{equation}
Thus the capacity of the G-WTC is the difference between the Shannon capacities of the main and eavesdropper's channels. We strengthen the main result in \cite{leung} as follows:
\begin{theorem}[Gaussian Wiretap Channels] \label{thm:str_conv_gauss}
The    (memoryless, stationary) G-WTC  satisfies the  partial strong converse under any secrecy metric  $\bbS_i,i \in [1:6]$.  Consequently, the $(i)$-capacities and $(i)$-optimistic capacities  of $\bW=\{W\}$ under the cost constraint in~\eqref{eqn:cost} are equal to $\CG(W;S)$ for all $i\in [1:6]$.
\end{theorem}
The proof of this theorem, which builds on that of Theorem~\ref{thm:str_conv_dwtc}, is provided in  Section~\ref{sec:prf_gauss}. 

One of the additional complications (vis-\`a-vis Theorem \ref{thm:str_conv_dwtc})  we have to overcome is the need to carefully handle the almost sure cost constraint in \eqref{eqn:cost} to ensure the statement holds for $\bbS_1$. Similarly to the proof of Theorem~\ref{thm:str_conv_gauss}, one can show, using the discretization procedure outlined in  Han {\em et al.}~\cite[Sec.~VI]{Han2013},  that the degraded Poisson wiretap channel, studied by Laourine and Wagner~\cite{Laourine}, admits a partial strong converse. 

\section{Conclusion and Future Work}\label{sec:concl}
In this paper, we proved  partial  strong converse theorems for various classes of degraded wiretap channels, including DM-WTCs and G-WTCs. We discuss three promising avenues for  further research.  

First, in this paper, we were only concerned with the transmission of a single message from the sender to the legitimate receiver. Csisz\'ar and K\"orner \cite{CsiszarKorner78} considered the broadcast channel with confidential messages  model in which two messages are to be sent, both to the legitimate receiver and only one to the eavesdropper. The eavesdropper's signal is to be asymptotically independent of the non-intended message. It may be possible to prove a partial strong converse in this multi-terminal system but we note that the information spectrum technique does not extend in a straightforward manner to show that discrete memoryless multi-terminal systems, such as the multiple-access channel \cite{Han98}, admit the strong converse thus new techniques  must be developed. 

Second,  as in \cite{TomTan13a}, it may be possible to use the techniques contained herein to study wiretap channels with limited memory (such as channels with additive Markov noise) and show that they admit a  partial strong converse.  However, wiretap channels with Markov memory have not been studied previously.

Finally, and most ambitiously, it would be interesting to study whether a {\em full}, and not partial or pretty~\cite{morgan}, strong converse holds for some classes of wiretap channels, i.e., whether the capacity depends on $(\eps,\delta)$ for  $\eps+\delta<1$. However, this appears to require general capacity formula with non-vanishing error probability and non-vanishing leakage, which in turn requires the evaluation a convenient non-asymptotic converse bound for channel resolvability. Initial work on refinements of non-asymptotic and asymptotic channel resolvability bounds has been conducted by Watanabe and Hayashi~\cite{WH14}. On a separate note, one-shot (non-asymptotic) bounds on the wiretap   capacity  for non-zero $(\eps,\delta)$ were proved by Renes and Renner~\cite{renes} using min- and max-entropy calculus. 

\section{Proofs} \label{sec:prfs}
\subsection{Proof Sketch of Theorem~\ref{prop:cap}}\label{sec:prf_cap}
\begin{proof}
We prove the achievability statement for the strongest secrecy metric $\bbS_2$ and the converse statement for the weakest secrecy metric $\bbS_6$. 

For achievability, fix a sequence of  input distributions $P_{\bV \bX}=\{P_{V^n X^n}\}_{n\in\bbN}$. For each message $m\in [1:M_n]$, generate a subcodebook $\calC(m)$ consisting of $\tilM_n / M_n$ randomly and independently generated sequences $\bv(l), l \in [1+ (m-1)\tilM_n / M_n  : m \tilM_n / M_n]$, each according to $P_{V^n}$. The codebook is revealed to all parties including the eavesdropper.  Given $m \in [1:M_n]$, the encoder chooses an index $L$ uniformly at random from  $[ 1+(m-1)\tilM_n / M_n : m \tilM_n / M_n]$ and generates $\bx(m) \sim P_{X^n| V^n}(\cdot | \bv( L))$ as the channel input. 

Let $\gamma>0$. Given $\by\in\calY^n$, the legitimate receiver finds the unique message $\hatm$ such that $(\bv(l), \by) \in \calT^{(n)}_\gamma$ for some $\bv(l) \in \calC(\hatm)$, where 
\begin{equation}
\calT^{(n)}_\gamma:=\left\{ (\bv,\by): \frac{1}{n}\log\frac{P_{Y^n|V^n}(\by|\bv)}{P_{Y^n}(\by)}\ge\frac{1}{n}\log \tilM_n+\gamma\right\}.\label{eqn:Tgamma}
\end{equation}
Let $\eps_n$ be the average error probability  of the    legitimate receiver (over the random message and the random code) given by \eqref{eqn:decoding_err}. By a standard calculation, we have 
\begin{equation}
\eps_n\le  P_{V^n Y^n} \big( (\calV^n\times\calY^n)\setminus\calT_\gamma^{(n)}\big)+\exp(-n\gamma).\label{eqn:fein}
\end{equation}
From $\eps$-capacity    \cite[Sec.~3.4]{Han10} and $\eps$-optimistic capacity analysis  \cite[Thm.~4.3]{chen99} (or simply by applying the definitions of $\calT^{(n)}_\gamma$,  $\underI_\eps$ and $\overI_\eps$ to~\eqref{eqn:fein}), we know that if $\tilM_n$ is chosen such that 
\begin{equation}
\frac{1}{n}\log \tilM_n  \le \underI_\eps (\bV;\bY) -2\gamma, \label{eqn:pess_m}
\end{equation}
then $\limsup_{n\to\infty}\,    \bbE[\eps_n]\le\eps$, where the expectation is over the random code. Similarly if $\tilM_n$ is chosen such that  
\begin{equation}
\frac{1}{n}\log \tilM_n   \le \overI_\eps (\bV;\bY) -2\gamma,\label{eqn:opt_m}
\end{equation}
then $\liminf_{n\to\infty}\,   \bbE[\eps_n] <\eps$. From the  {\em secrecy from resolvability} condition  in \cite[Lem.~2]{Bloch}, we know that if 
\begin{equation}
\frac{1}{n}\log\tilM_n-\frac{1}{n}\log M_n \ge \overI(\bV;\bZ) + 2\gamma \label{eqn:sec_from_res}
\end{equation}
then $\lim_{n\to\infty}\bbE[\bbS_2] =0$. Now because averaged over the random code, $\bbS_2$ tends to zero,   by a Markov inequality argument (see proof of~\cite[Thm.~1]{TanMoulin14} for example), there exists  a sequence of codes such that both the reliability and security conditions are satisfied. This completes the direct part of Theorem~\ref{prop:cap} upon eliminating $\tilM_n$ from the above inequalities, taking $\liminf_{n\to\infty}$, and finally taking $\gamma\downarrow 0$. 

For the converse, by using the Verd\'u-Han lemma~\cite[Lem.~4]{VH94} we know that if  $\limsup_{n\to\infty}\,    \eps_n \le\eps$, for every $\gamma>0$, we must have that  
\begin{equation}
\frac{1}{n}\log M_n  \le \underI_\eps (\bV;\bY) + \gamma,  \label{eqn:pess2}
\end{equation}
for some chain $\bV-\bX-(\bY,\bZ)$ and all $n$ sufficiently large (depending on $\gamma$).   The auxiliary random process $\bV$ represents the  sequence of messages  which are uniform on the message sets $\{\calM_n\}_{n\in\bbN}$. Similarly, if   $\liminf_{n\to\infty}\,    \eps_n < \eps$, we must have that 
\begin{equation}
\frac{1}{n}\log M_n   \le \overI_\eps (\bV;\bY)  +  \gamma. \label{eqn:opt2}
\end{equation}
Furthermore,  \cite[Lem.~4]{bloch08} tells  us that if $\bbS_6\to 0$, we must have that 
\begin{equation}
\overI(\bV;\bZ)=0.
\end{equation}
This follows directly from the definition  of $\bbS_6$ in \eqref{eqn:S6} and the spectral sup-mutual information rate. 
Subtracting $\overI(\bV;\bZ)$ from \eqref{eqn:pess2} and \eqref{eqn:opt2}, maximizing over all chains $\bV-\bX-(\bY,\bZ)$ to  make the bound code-independent, and finally taking  $\liminf_{n\to\infty}$ and $\gamma\downarrow 0$ completes  the converse proof of Theorem~\ref{prop:cap}. 
\end{proof}
\subsection{Proof of Theorem~\ref{thm:str_conv_dwtc}} \label{sec:prf_degraded}
\begin{proof}
Here we prove that any degraded DM-WTC $W:\calX\to\calY\times\calZ$ satisfies the partial strong converse for any secrecy metric $i \in [1:6]$. We proceed in two steps. First, we show that $C^{(1)}(\bW) \ge \CDM(W)$ (where $\bW$ is the stationary, memoryless channel induced by $W$) and second, we show that $\overC^{(6)}(\bW) \le \CDM(W)$. 

To show that $C^{(1)}(\bW) \ge \CDM(W)$, we adopt the strategy in \cite[Sec.~V.C]{Bloch}. Particularize the supremum over  $P_{\bV \bX}$  by choosing $\bV=\bX$ and  $P_{\bX}$ to be a sequence of product distributions induced by any $P_{\barX} \in\argmax_{P_X\in\calP(\calX)}I(X;Y|Z)$. Then, it  suffices to appeal to  \cite[Rmk.~3]{Bloch} which says that if 
\begin{equation}
\fp_n:=\Pr\bigg( \frac{1}{n}\log\frac{W^n_{\calZ}(Z^n|X^n)}{ P_{Z^n}(Z^n)} \ge\frac{1}{n}\log \frac{\tilM_n}{M_n} -\gamma\bigg) \label{eqn:pn}
\end{equation}
decays exponentially   in $n$ then  $\bbS_1 \to 0$. This remark was also made by Kobayashi {\em et al.}~\cite[Sec.~V]{KYO13}, and is a simple consequence of a bound presented by Csisz\'ar in \cite[Lem.~1]{csi96} relating mutual information to variational distance.  Choose  $\tilM_n$ to be the smallest integer exceeding $\exp[n  (I(X;Y )-2\gamma)  ]$ (so the decoding error probability tends to zero),  and choose $M_n$ to be the largest integer smaller than $\exp[n( I(X;Y ) - I(X;Z)-4\gamma ) ] = \exp[n(\CDM(W)-4\gamma)]$. The mutual informations are computed with respect to the distribution $P_{\barX}\times W$. Now,   we see that \eqref{eqn:pn} indeed decays exponentially (Chernoff bound) and so $C^{(1)}(\bW) \ge \CDM(W) - 4\gamma$. Finally, let $\gamma\downarrow 0$.

Now, we prove that $\overC^{(6)}(\bW)\le\CDM(W)$. Starting from~\eqref{eqn:C6}, for every  Markov chain $\bV -\bX-(\bY,\bZ)$, we  have 
\begin{align}
\overI  (\bV;\bY) - \overI(\bV;\bZ) \le \overI  (\bV;\bY,\bZ) - \overI(\bV;\bZ)\le \overI(\bV;\bY|\bZ) \label{eqn:def_plims2}
\end{align}  
where the final inequality follows from the  sub-additivity of $\plimsup$ \cite[Sec.~1.3]{Han10}, i.e., that  
\begin{equation}
\plimsup_{n\to\infty}(A_n+B_n)  \le\plimsup_{n\to\infty} A_n + \plimsup_{n\to\infty} B_n.\label{eqn:plimsup_subadd1}
\end{equation}
We further upper bound $\overI(\bV;\bY|\bZ)$ in \eqref{eqn:def_plims2}. By a conditional version of the data processing inequality    \cite[Thm.~9]{VH94},  
\begin{equation}
\overI(\bV;\bY|\bZ)\le \overI(\bX;\bY|\bZ)\label{eqn:MbyX}
\end{equation}
because $\bV - \bX - (\bY,\bZ)$ forms  a Markov chain. Thus,  
\begin{equation}
  \overC^{(6)}(\bW)\le\sup_{\bX}  \overI(\bX;\bY|\bZ) \label{eqn:spectral_sup}
\end{equation}
for any general wiretap channel $\bW=\{W^n\}_{n\in\bbN}$.  Now, it suffices to simplify the spectral sup-conditional mutual information rate in \eqref{eqn:spectral_sup} and, in particular, to show that 
\begin{equation}
\sup_{\bX}  \overI(\bX;\bY|\bZ)\le \CDM(W) , \label{eqn:opt_sup_cmir1}
\end{equation}
where $\CDM(W)$ is the capacity of the degraded DM-WTC defined in \eqref{eqn:cap_dmtc}.
At this point, we  note that Koga and Sato~\cite{koga} argued  (without proof) that $\sup_{\bX} \underI(\bX;\bY|\bZ)\le \CDM(W)$ for degraded DM-WTCs, but   \eqref{eqn:opt_sup_cmir1} is stronger as we optimize the spectral sup- (instead of the spectral inf-) conditional mutual information rate. Hence, an immediate corollary of~\eqref{eqn:opt_sup_cmir1} is Koga and Sato's claim~\cite{koga}. For this purpose, define  the conditional channel
\begin{equation}
W_{\calY|\calZ}(y|x,z) := \frac{W(y,z|x)}{\sum_{y \in\calY}W(y,z|x) }.
\end{equation}
We proceed to show \eqref{eqn:opt_sup_cmir1} by first considering the sequence of    random variables
\begin{equation}
\imath_n(X^n;Y^n|Z^n) := \frac{1}{n}\log\frac{W_{\calY|\calZ}^n(Y^n|X^n, Z^n)}{P_{Y^n|Z^n}(Y^n|Z^n)}
\end{equation}
where $\bX = \{X^n\}_{n\in\bbN}$ is an arbitrary input that induces the output random variables $(\bY,\bZ) = \{(Y^n,Z^n)\}_{n\in\bbN}$. Let $P_{\barY \barZ} \in\calP(\calY\times\calZ)$ be a single-letter capacity-achieving output distribution, i.e., a distribution on $\calY\times\calZ$ such that 
\begin{equation}
P_{\barY \barZ}(y,z):=\sum_{x\in\calX} P_{\barX}(x)W(y,z|x)  \label{eqn:Ybar_Zbar}
\end{equation}
for some $P_{\barX}\in\calP(\calX)$ that achieves the $\max$ in \eqref{eqn:cap_dmtc}.  By the same logic as \cite[Cor.~2 to Thm.~4.5.2]{gallagerIT},  $P_{\barY|\barZ}$ is unique. In contrast, $P_{\barZ}$ is not necessarily unique but, as we will see, this is inconsequential for the subsequent derivations.

For simplicity in notation, define
\begin{equation}
 \jmath(\bX)  :=\plimsup_{n\to\infty}\imath_n (X^n;Y^n|Z^n),
\end{equation}
where $\bX$ is an arbitrary input process.  Since the $\plimsup$ is sub-additive as in~\eqref{eqn:plimsup_subadd1}, by introducing the product distribution $P_{\barY |\barZ }^n$,  we obtain 
\begin{align}
\jmath(\bX)& =\plimsup_{n\to\infty} \Bigg(\frac{1}{n}\log\frac{W_{\calY|\calZ}^n(Y^n|X^n, Z^n)}{P_{\barY |\barZ }^n(Y^n|Z^n)}  \nn\\*
&\qquad\qquad\qquad\qquad - \frac{1}{n}\log\frac{P_{Y^n|Z^n}(Y^n| Z^n)}{P_{\barY |\barZ }^n(Y^n|Z^n)}  \Bigg)\\
&\le \plimsup_{n\to\infty} \frac{1}{n}\log\frac{W_{\calY|\calZ}^n(Y^n|X^n, Z^n)}{P_{\barY |\barZ }^n(Y^n|Z^n)}  \nn\\*
&\qquad\qquad - \pliminf_{n\to\infty}\frac{1}{n}\log\frac{P_{Y^n|Z^n}(Y^n|Z^n) }{P_{\barY |\barZ }^n(Y^n|Z^n)} . \label{eqn:pliminf_nonnega}
\end{align}
The final term is non-negative following~\cite[Lem.~3.2.1]{Han10} and hence 
\begin{align}
\jmath(\bX)\le \plimsup_{n\to\infty} \frac{1}{n}\log\frac{W^n_{\calY|\calZ}(Y^n|X^n, Z^n)}{P_{\barY |\barZ }^n(Y^n|Z^n)} . \label{eqn:nonneg}
\end{align}
Now let $X^n = (X^{(n)}_1, \ldots, X^{(n)}_n)$, $Y^n = (Y^{(n)}_1, \ldots, Y^{(n)}_n)$ and $Z^n = (Z^{(n)}_1, \ldots, Z^{(n)}_n)$ for each blocklength $n \in\bbN$. 
Since the channel $W^n$ and the conditional capacity-achieving  output measure $P_{\barY |\barZ }^n$ are memoryless, 
\begin{align}
\jmath(\bX) \le \plimsup_{n\to\infty} \frac{1}{n}\sum_{i=1}^n L_i^{(n)}(X_i^{(n)}) \label{eqn:removed}
\end{align}
where the information density random variables $L_i^{(n)}(x_i)$ are defined as 
\begin{align}
L_i^{(n)}(x_i):=\log\frac{W_{\calY|\calZ} (Y_i^{(n)}|x_i, Z_i^{(n)})}{P_{\barY |\barZ } (Y_i^{(n)}|Z_i^{(n)})}. \label{eqn:removed2}
\end{align}
Now by a result of  Yasui {\em et al.}~\cite[Lem.~1]{yasui}, we know that  for every $x\in\calX$,
\begin{equation}
\bbE\left[\log\frac{W_{\calY|\calZ}(Y|x,Z)}{P_{\barY|\barZ}(Y|Z)}\right]\le \CDM(W) \label{eqn:info_radiu}
\end{equation}
where $(Y,Z) |\{X=x\} \sim W(\cdot,\cdot|x)$.    This follows from the KKT conditions and straightforward differentiation of mutual information with respect to the input distribution (cf.~\cite[Thm.~4.5.1]{gallagerIT}). 
Note that we used the fact that $W$ is degraded to establish \eqref{eqn:info_radiu}. From \eqref{eqn:info_radiu} and the definition of $L_i^{(n)}(x_i)$ in~\eqref{eqn:removed2}, for every $\bx = (x_1, \ldots, x_n) \in\calX^n$, we have 
\begin{equation}
\bbE\left[ \frac{1}{n}\sum_{i=1}^n L_i^{(n)}(x_i)\right]\le \CDM(W). \label{eqn:bound_expect}
\end{equation}
Because we fixed a deterministic $\bx$ and the channel is memoryless, the   random variables $(Y_i^{(n)}, Z_i^{(n)}), i = 1,\ldots, n$  are independent under the channel    $W^n(\cdot,\cdot|\bx)$.  
By memorylessness and  Chebyshev's inequality, for every $\gamma>0$,
\begin{align}
 \Pr\left(  \frac{1}{n}\sum_{i=1}^n L_i^{(n)}(x_i)\!\ge\! \CDM(W) \!+\! \gamma    \bigg|     X^n \!=\!\bx \right) \!\le\!\frac{\sigma_0^2}{n\gamma^2}, \label{eqn:det_x}
\end{align}
where the constant $\sigma_0^2$ is defined as 
\begin{equation}
\sigma_0^2 :=\max_{x\in\calX} \var\left[\log \frac{W_{\calY|\calZ}(Y | x, Z  )}{P_{\barY|\barZ}(Y |Z )} \right]. \label{eqn:variance}
\end{equation}
The constant $\sigma_0^2$ is finite because $P_{\barY|\barZ}(y| z)$ is positive  for all $(y,z)$ in view of \eqref{eqn:info_radiu} and the finiteness of $\CDM(W)\le\min\{\log|\calX|,\log|\calY|\}$. Since \eqref{eqn:det_x} is true  uniformly over every $\bx\in\calX^n$, we may average it over $\bx$ to obtain 
\begin{align}
\Pr\left(  \frac{1}{n}\sum_{i=1}^n L_i^{(n)}(X_i^{(n)} )\ge \CDM(W) + \gamma \right)   \le \frac{\sigma_0^2}{n\gamma^2}.
\end{align}
The upper bound $\sigma_0^2 / (n\gamma^2)$ clearly tends to zero as $n\to\infty$. From the definition of $\plimsup$ and $L_i^{(n)}(X_i^{(n)})$,  we have 
\begin{equation}
\plimsup_{n\to\infty} \!\frac{1}{n}\sum_{i=1}^n\! \log  \frac{W_{\calY|\calZ}(Y_i^{(n)} | X_i^{(n)}, Z_i^{(n)} )}{P_{\barY|\barZ}(Y_i^{(n)}|Z_i^{(n)})}\!\le\! \CDM(W)\!+\! \gamma.
\end{equation}
Consequently, from \eqref{eqn:nonneg} and \eqref{eqn:removed}, this proves that 
\begin{equation}
\sup_{\bX}\overI(\bX;\bY|\bZ)\le \CDM(W)+\gamma.
\end{equation}
Since $\gamma$ is arbitrary,  we may take $\gamma\downarrow 0$. That is, we  have proved the claim in \eqref{eqn:opt_sup_cmir1}, completing the proof of the partial strong converse for degraded DM-WTCs.
\end{proof}

\subsection{Proof of Theorem~\ref{thm:str_conv_non}} \label{sec:prf_non_stat}
\begin{proof}
We assume that the limits in \eqref{eqn:cesaro_means} exist. We will prove that 
\begin{equation}
C^{(1)}(\bW)\ge \liminf_{n\to\infty}\,   \frac{1}{n}\sum_{i=1}^n C(W_{\calY,i})-\limsup_{n\to\infty}\,   \frac{1}{n}\sum_{i=1}^n C(W_{\calZ,i})  \label{eqn:est1}
\end{equation}
as well as 
\begin{equation}
\overC^{(6)}(\bW)\le \limsup_{n\to\infty}\,   \frac{1}{n}\sum_{i=1}^n \CDM(W_i). \label{eqn:est2}
\end{equation}
Indeed, if  the limits in \eqref{eqn:cesaro_means} exist, it is easy  to see from \eqref{eqn:diff_cap} that the right-hand-sides of \eqref{eqn:est1} and \eqref{eqn:est2} are equal and thus  $C^{(1)}(\bW)=\overC^{(6)}(\bW)$. This implies that $C^{(i)}(\bW)=\overC^{(i)}(\bW)$ for all $i\in [1:6]$, i.e., the partial strong converse holds for secrecy metrics $\bbS_i,i\in [1:6]$.

For inequality \eqref{eqn:est1}, we first show the weaker statement:
\begin{equation}
C^{(2)}(\bW)\ge \liminf_{n\to\infty}\,   \frac{1}{n}\sum_{i=1}^n C(W_{\calY,i})-\limsup_{n\to\infty}\,   \frac{1}{n}\sum_{i=1}^n C(W_{\calZ,i})  \label{eqn:est1_C2} .
\end{equation}
For this purpose, we follow the steps in the proof of~\cite[Cor.~3]{Tan14} for the non-stationary Gel'fand-Pinsker channel. Particularize the optimization over $\bV-\bX-(\bY,\bZ)$ to $\bV=\bX$ being  uniform on $\calX^n$ for every $n \in\bbN$. Invoking Theorem~\ref{prop:cap}, we then find
\begin{align}
&C^{(2)}(\bW) \nn\\*
&\ge  \pliminf_{n\to\infty}\frac{1}{n} \log \frac{ W_\calY^n(Y^n | X^n)}{ P_{Y^n}(Y^n) }  \nn\\*
&\qquad- \plimsup_{n\to\infty}\frac{1}{n}\log \frac{ W_\calZ^n(Z^n | X^n)}{ P_{Z^n}(Z^n) } \\
&=   \pliminf_{n\to\infty}\frac{1}{n} \sum_{i=1}^n \log \frac{ W_{\calY,i}(Y_i | X_i )}{ P_{Y_i}(Y_i) } \nn\\* 
&\qquad - \plimsup_{n\to\infty}\frac{1}{n}\sum_{i=1}^n\log \frac{ W_{\calZ,i}(Z_i | X_i)}{ P_{Z_i}(Z_i) }  \label{eqn:memo}\\
&=   \liminf_{n\to\infty}\,  \frac{1}{n} \sum_{i=1}^n C(W_{\calY,i})- \limsup_{n\to\infty}\,  \frac{1}{n}\sum_{i=1}^n C(W_{\calZ,i}) \label{eqn:cheb1}
\end{align}
where \eqref{eqn:memo} follows from memorylessness and~\eqref{eqn:cheb1} follows from Chebyshev's inequality and the fact that the alphabets are finite. See  \cite[Eq.~(3.2.15)]{Han10} for a similar statement. 

Now, we prove the stronger statement in~\eqref{eqn:est1} concerning $C^{(1)}(\bW)$. Given we have proved \eqref{eqn:est1_C2},  it suffices \cite[Rmk.~3]{Bloch} to  verify that   $\fp_n$ in \eqref{eqn:pn} (which controls the variational distance) converges to zero with rate $O(1/n^2)$. This is because according to \cite[Lem.~1]{csi96}, 
\begin{equation}
I(M;Z^n)\le \bbV(P_{MZ^n}, P_M\times P_{Z^n}) \log \frac{|\calM_n|}{\bbV(P_{MZ^n}, P_M\times P_{Z^n})}.
\end{equation}
Since $\log  |\calM_n| $ is linear in $n$ (cf.~\eqref{eqn:pess_def} and \eqref{eqn:opt_def1}),   if the variational distance  $\bbV(P_{MZ^n}, P_M\times P_{Z^n})$ decays as   $O(1/n^2)$, the mutual information $I(M;Z^n)$ decays as  $O(1/n)$. Choose $\bX=\{X^n\}_{n\in\bbN}$ such that $X^n$ is uniform on $\calX^n$ for each $n$.  In addition, if  we choose $\tilM_n$ in \eqref{eqn:pn} to be the smallest  integer exceeding $\exp[n  (\underI(\bX;\bY )-2\gamma)  ]$,  and   $M_n$   to be the largest integer smaller than $\exp[n( \underI(\bX;\bY ) - \overI(\bX;\bZ)-4\gamma ) ]$, we have 
\begin{equation}
\fp_n\le\Pr\bigg( \frac{1}{n}\log\frac{W^n_{\calZ}(Z^n|X^n)}{ P_{Z^n}(Z^n)} \ge\overI(\bX;\bZ)+\gamma\bigg)  \label{eqn:pn_non_stat} .
\end{equation}
Furthermore,  by the same argument  that led to~\eqref{eqn:cheb1},  we notice that  with $X^n$ uniform on $\calX^n$,
\begin{equation}
\overI(\bX;\bZ) =\limsup_{n\to\infty}\, \frac{1}{n}\sum_{i=1}^n C(W_{\calZ,i}).
\end{equation}
Thus, for every $\gamma>0$,  there exists an integer $N_\gamma'$ such that for all $n>N_\gamma'$,  
\begin{equation}
\overI(\bX;\bZ)+\frac{\gamma}{2}\ge\frac{1}{n}\sum_{i=1}^n C(W_{\calZ,i}).\label{eqn:pn_non_stat2}
\end{equation}
Uniting \eqref{eqn:pn_non_stat} and \eqref{eqn:pn_non_stat2} and invoking the memorylessness of $W_{\calZ}^n$,  we have 
\begin{equation}
\fp_n\le\Pr\bigg( \frac{1}{n}\sum_{i=1}^n\Big(\log\frac{W_{\calZ,i}(Z_i|X_i)}{ P_{Z_i}(Z_i)}  - C(W_{\calZ,i}) \Big)\ge \frac{\gamma}{2}\bigg).
\end{equation}
To prove that $\fp_n=O(1/n^2)$, we use a similar proof strategy as the strong law of large numbers assuming finite fourth moments (e.g., \cite[Thm.~2.3.5]{Durrett}). To simplify notation,    define the zero-mean, independent (but not identically distributed) random variables 
\begin{equation}
 J_i := \log\frac{W_{\calZ,i}(Z_i|X_i)}{ P_{Z_i}(Z_i)}  - C(W_{\calZ,i}) . \label{eqn:defJi} 
\end{equation} 
Then by Markov's inequality, we have 
\begin{align}
\fp_n  \le \Pr\left( \Big(\frac{1}{n}\sum_{i=1}^n J_i\Big)^4 \ge \frac{\gamma^4}{16}\right) \le \frac{16}{n^4\gamma^4} \cdot  \bbE \left[ \Big( \sum_{i=1}^n J_i\Big)^4 \right]. \label{eqn:markov}
\end{align}
When we expand $(\sum_{i=1}^n J_i)^4$ and take expectation, the only terms with $\bbE[J_{i_1}J_{i_2}J_{i_3}J_{i_4}]\ne 0$ are the ones where $i_1,\ldots, i_4$ are all equal, or they take on two distinct values with each value repeated twice among $i_1,\ldots, i_4$. In other words, 
\begin{equation}
 \bbE \left[ \Big( \sum_{i=1}^n J_i\Big)^4 \right] = \sum_{i=1}^n \bbE[J_i^4] + 6 \sum_{1\le i<j\le n }\bbE[J_i^2]\bbE[J_j^2].
\end{equation}
Now we must argue that each of the terms $\bbE[J_i^k]$ for $k=2,4$ is uniformly bounded in $i$ (but possibly dependent on $|\calX|$ and $|\calZ|$). Then, because of the normalization by $n^4$ in \eqref{eqn:markov}, we have the desired convergence rate of $\fp_n$. Indeed, it is easy to see that for this assertion to be true, it suffices to show that the first four moments of the information density random variable $\log W_{\calZ,i}(Z_i|X_i)-\log P_{Z_i}(Z_i)$ are bounded (since the capacity terms in \eqref{eqn:defJi} are uniformly bounded). Now, note that $P_{Z_i}(z)=1/|\calZ|$ for all $z\in\calZ$  by the weak symmetry of the channels. Hence, it suffices to show that $\bbE [ (\log W_{\calZ,i}(Z_i|X_i) + \log|\calZ|)^k   ]$  are uniformly bounded for each $k\in [1:4]$. However, it then suffices to verify that $\bbE [  \log^k W_{\calZ,i}(Z_i|X_i)]$ are uniformly bounded. This immediately follows from the fact that $u\in [0,1]\mapsto |u \log^k u|$   is bounded above by $\rme^{-k}k^k$ (assuming natural logs).


Now we prove inequality \eqref{eqn:est2}. By using \eqref{eqn:info_radiu}, we know that for every $\bx\in\calX^n$,
\begin{equation}
\bbE\left[ \frac{1}{n}\sum_{i=1}^n L_i^{(n)}(x_i)\right]\le \frac{1}{n}\sum_{i=1}^n\CDM(W_i),   \label{eqn:expect_sup}
\end{equation}
where  the random variable
\begin{equation}
L_i^{(n)}(x_i) := \log \frac{W_{\calY|\calZ,i}(Y_i^{(n)} | x_i, Z_i^{(n)} )}{P_{\barY|\barZ,i}(Y_i^{(n)}|Z_i^{(n)})}
\end{equation}
and $W_{\calY|\calZ,i} : \calX\times\calZ\to\calY$ and $P_{\barY|\barZ,i}: \calZ\to\calY$ are induced by $W_{i}: \calX\to\calY\times\calZ$. Note that we leveraged  the degradedness of the channels $\{W_i\}_{i\in\bbN}$ to arrive at~\eqref{eqn:expect_sup}.  Define
\begin{equation}
C^\ddagger:=\limsup_{n\to\infty}\,   \frac{1}{n}\sum_{i=1}^n \CDM(W_i).
\end{equation}
 By the definition of $\limsup$, for every $\gamma>0$, there exists an integer $N_\gamma$ such that for all $n>N_\gamma$, we have 
\begin{equation}
\frac{1}{n}\sum_{i=1}^n \CDM(W_i)\le  C^\ddagger+\gamma. \label{eqn:def_limsup}
\end{equation}
Uniting \eqref{eqn:expect_sup} and \eqref{eqn:def_limsup}, we obtain
\begin{equation}
\bbE\left[ \frac{1}{n}\sum_{i=1}^n L_i^{(n)}(x_i) \right]\le C^\ddagger+\gamma, \label{eqn:dagger_bd}
\end{equation}
for all $n>N_\gamma$. Let $\sigma_0^2$, analogously to \eqref{eqn:variance}, be defined as 
\begin{equation}
\sigma_0^2 := \sup_{i\in\bbN}\max_{x\in\calX} \var\left[\log \frac{W_{\calY|\calZ,i}(Y | x, Z  )}{P_{\barY|\barZ,i}(Y |Z )} \right]. \label{eqn:variance2}
\end{equation}
We would now like to show that $\sigma_0^2$ is finite.
By Bayes rule and the degradedness of each channel $W_i$,
\begin{align}
\log \frac{W_{\calY|\calZ,i}(Y | x, Z  )}{P_{\barY|\barZ,i}(Y |Z )} 
  = \log \frac{W_{\calY,i}(Y | x   )}{P_{\barY ,i}(Y  )}  - \log \frac{W_{\calZ,i}(Z | x   )}{P_{\barZ ,i}(Z  )} .
\end{align}
By using the fact that $\var[A+B]\le 2\var[A] + 2\var[B]$,  it is enough to show that 
\begin{equation}
\var\left[\log \frac{W_{\calY,i}(Y | x   )}{P_{\barY ,i}(Y  )}\right] ,\,\,\, \mbox{and}\,\,\, \var\left[ \log \frac{W_{\calZ,i}(Z | x   )}{P_{\barZ ,i}(Z  )} \right] \label{eqn:XY}
\end{equation}
are uniformly bounded in  $i\in\bbN$.  Now note that $P_{\barY,i}$ and $P_{\barZ,i}$ are uniform on $\calY$ and $\calZ$ respectively (by the symmetry of the channels) so 
\begin{align}
\var\left[\log \frac{W_{\calY,i}(Y | x   )}{P_{\barY ,i}(Y  )}\right] & = \var\big[\log W_{\calY,i}(Y | x   )  \big] \\
& \le \bbE\left[\log^2 W_{\calY,i}(Y | x   )  \right] \\
& \le 4\rme^2 \cdot |\calY|
\end{align}
where the last inequality follows from the fact that $\sup_{u\in (0,1]} |u \log^2 u|\le  4\rme^2$.  A similar calculation can be done for the second term in \eqref{eqn:XY}. Thus $\sigma_0^2$ is finite. 

By Chebyshev's inequality and~\eqref{eqn:dagger_bd} (the same logic that led to \eqref{eqn:det_x}),  we have 
\begin{equation}
\Pr\left( \frac{1}{n}\sum_{i=1}^n L_i^{(n)}(x_i) \ge C^\ddagger+2\gamma \, \bigg|\, X^n=\bx\right) \le\frac{\sigma_0^2}{n\gamma^2} , \label{eqn:det_x2}
\end{equation}
for all $n>N_\gamma$ and all $\bx\in\calX^n$.  It is also true that
\begin{equation}
\Pr\left( \frac{1}{n}\sum_{i=1}^n L_i^{(n)}(X_i^{(n)}) \ge C^\ddagger+2\gamma \right) \le\frac{\sigma_0^2}{n\gamma^2} , \label{eqn:det_x3}
\end{equation}
holds for all $n>N_\gamma$. 
By the definition of $\plimsup$ and $L_i^{(n)}(X_i^{(n)})$, 
\begin{equation}
\plimsup_{n\to\infty}\frac{1}{n}\sum_{i=1}^n \log  \frac{W_{\calY|\calZ,i}(Y_i^{(n)} | X_i^{(n)}, Z_i^{(n)} )}{P_{\barY|\barZ,i}(Y_i^{(n)}|Z_i^{(n)})}\le C^\ddagger+2\gamma.
\end{equation}
Finally, from \eqref{eqn:nonneg}, we have 
\begin{equation}
\overC^{(6)}(\bW)\le\sup_{\bX}\overI(\bX;\bY|\bZ)\le C^\ddagger+2\gamma.
\end{equation}
Since this holds for all $\gamma>0$, we may take $\gamma\downarrow 0$ to complete the proof of \eqref{eqn:est2}.
\end{proof}
\subsection{Proof of Theorem~\ref{thm:str_conv_gauss}} \label{sec:prf_gauss}
\begin{proof}
Similarly to the proof of Theorem~\ref{thm:str_conv_dwtc}, we show that $C^{(1)}(\bW) \ge \CG(W;S)$ and $\overC^{(6)}(\bW) \le \CG(W;S)$. Note,  however, that  the form of the optimistic capacity $\overC^{(6)}(\bW)$ in \eqref{eqn:C6}  has to be modified to take into account the cost constraint $\Pr(X^n\in\calF_n)=1$ in \eqref{eqn:cost}. The optimization over the chain $\bV-\bX-(\bY,\bZ)$ has to be further constrained to all distributions $P_{\bV\bX}$ satisfying $X^n\in\calF_n$ for all $n\in\bbN$.

For the lower bound, $C^{(1)}(\bW) \ge \CG(W;S)$, we need to show that  $\fp_n$, defined in \eqref{eqn:pn}, decays exponentially fast for an appropriate choice of input distribution. This argument is adapted from  the proofs of Lemmas 2 and 5 in He and Yener~\cite{HeYener}. Fix a constant $\delta>0$ and define  the product distribution (probability density function)
\begin{equation}
  P_{\tilX^n}(\bx) := \prod_{i=1}^n\calN(x_i ; 0, S-\delta). \label{eqn:prod_dist}
\end{equation}  
Now define the input distribution to be 
\begin{equation}
P_{X^n}(\bx) := \frac{ P_{\tilX^n}(\bx)  }{\mu_n } \bone\{\bx\in\calF_n\}\label{eqn:non_prod_d}
\end{equation}
where $\mu_n$ is the normalizing constant that ensures that $\int P_{X^n}(\bx)\,\rmd\bx=1$. This is simply a truncated version of the jointly Gaussian distribution $P_{\tilX^n}$ in \eqref{eqn:prod_dist}.  Because of the constant backoff $\delta>0$ from the permissible power $S$ in~\eqref{eqn:prod_dist}, it can be seen from Cramer's large deviations theorem on the real line~\cite[Sec.~2.2]{Dembo} that $\mu_n := P_{\tilX^n}(\calF_n)$ tends to $1$ exponentially fast, i.e., 
\begin{equation}
 \mu_n = \Pr\bigg( \frac{1}{n}\sum_{i=1}^n \tilX_i^2\le S\bigg)\ge 1-\exp(- n\eta_1) \label{eqn:mu_behave}
\end{equation}
 for some $\eta_1>0$ depending on $\delta$.  By  the construction of the input distribution in \eqref{eqn:non_prod_d}, $X^n \in\calF_n$ with probability one, satisfying the almost sure power constraint in \eqref{eqn:cost}. Using the characterization of the variational distance in~\eqref{eqn:var_dist2}, we have 
\begin{align}
&\bbV(P_{\tilX^n}, P_{X^n}) \nn\\*
& = \frac12\int_{\bbR^n} \big| P_{\tilX^n}(\bx) -  P_{X^n}(\bx) \big|\, \rmd \bx\\
& = \frac12\int_{\calF_n} \big| P_{\tilX^n}(\bx) -  P_{X^n}(\bx) \big|\, \rmd\bx  \nn\\*
&\qquad +  \frac12 \int_{\calF_n^c} \big| P_{\tilX^n}(\bx) -  P_{X^n}(\bx) \big|\, \rmd \bx\\
& = \frac12\int_{\calF_n}   P_{X^n}(\bx)\big|\mu_n -  1 \big|\, \rmd\bx  +   \frac12P_{\tilX^n}( \calF_n^c ) \label{eqn:zero}\\
&\le \frac12\exp(-n \eta_1) +  \frac12\exp(-n \eta_1)  \\
&=\exp(-n\eta_1)\label{eqn:bounds}
\end{align}
where \eqref{eqn:zero} follows from the definition of $P_{X^n}(\bx)$, and \eqref{eqn:bounds} follows from \eqref{eqn:mu_behave}. Consequently,
\begin{align}
\bbV(P_{\tilX^n}\times W_\calZ^n, P_{X^n}\times W_\calZ^n)&=\bbV(P_{\tilX^n}, P_{X^n}) \\
&\le \exp(-n \eta_1). \label{eqn:exp_small_jt}
\end{align}
Let $(\tilX,\tilZ)\sim P_{\tilX_1}\times W_\calZ$, $(\tilX^n,\tilZ^n)\sim P_{\tilX^n}\times W_\calZ^n$ and  $(X^n,Z^n)\sim P_{X^n}\times W_\calZ^n$. By using the characterization of the variational distance in \eqref{eqn:var_dist} as well as the bound in~\eqref{eqn:exp_small_jt}, we deduce that for any $\beta\in\bbR$, 
\begin{align}
&\bigg| \Pr\bigg( \frac{1}{n}\log\frac{W^n_{\calZ}(Z^n|X^n)}{ P_{\tilZ^n}(Z^n)} \ge\beta\bigg)   \nn\\*
&\qquad  - \Pr\bigg( \frac{1}{n}\log\frac{W^n_{\calZ}(\tilZ^n|\tilX^n)}{ P_{\tilZ^n}(\tilZ^n)} \ge\beta\bigg) \bigg| \le \exp(-n \eta_1). \label{eqn:var_joint}
\end{align}
Define $\alpha :=\frac{1}{n}\log (\tilM_n/M_n) -\gamma$. Let $\eta_2>0$ be an arbitrary constant for now. The probability   $\fp_n$ in \eqref{eqn:pn}   can be  written and bounded as 
\begin{align}
  \fp_n  &= \Pr\bigg( \frac{1}{n}\log\frac{W^n_{\calZ}(Z^n|X^n)}{ P_{Z^n}(Z^n)} \ge\alpha\bigg) \\
  &= \Pr\bigg( \frac{1}{n}\log\frac{W^n_{\calZ}(Z^n|X^n)}{ P_{\tilZ^n}(Z^n)}-\frac{1}{n}\log \frac{P_{Z^n}(Z^n)}{P_{\tilZ^n}(Z^n)} \ge\alpha\bigg) \\
    &\le \Pr\bigg( \frac{1}{n}\log\frac{W^n_{\calZ}(Z^n|X^n)}{ P_{\tilZ^n}(Z^n)}-\frac{1}{n}\log \frac{P_{Z^n}(Z^n)}{P_{\tilZ^n}(Z^n)} \ge\alpha  \nn\\*
    &\qquad\qquad \mbox{and}\,\,\,\, \frac{1}{n}\log \frac{P_{Z^n}(Z^n)}{P_{\tilZ^n}(Z^n)} >-\eta_2\bigg) \nn\\*
     &\qquad \qquad\qquad+\Pr\bigg( \frac{1}{n}\log \frac{P_{Z^n}(Z^n)}{P_{\tilZ^n}(Z^n)} \le-\eta_2\bigg) \\
      &\le \Pr\bigg( \frac{1}{n}\log\frac{W^n_{\calZ}(Z^n|X^n)}{ P_{\tilZ^n}(Z^n)}  \ge\alpha  -\eta_2\bigg)+\exp(-n\eta_2) \\
       &\le \Pr\bigg( \frac{1}{n}\log\frac{W^n_{\calZ}(\tilZ^n|\tilX^n)}{ P_{\tilZ^n}(\tilZ^n)}  \ge\alpha  -\eta_2\bigg) \nn\\*
       &\qquad\qquad\qquad+\exp(-n\eta_2) +\exp(-n\eta_1)\label{eqn:use_var} \\
        &=\Pr\bigg( \frac{1}{n}\sum_{i=1}^n\log\frac{W_{\calZ}(\tilZ_i|\tilX_i)}{ P_{\tilZ }(\tilZ_i)}  \ge\alpha  -\eta_2\bigg) \nn\\*
        &\qquad\qquad\qquad+\exp(-n\eta_2) +\exp(-n\eta_1) \label{eqn:use_iid}
\end{align}
where \eqref{eqn:use_var} uses the bound in \eqref{eqn:var_joint} with the identification $\beta=\alpha-\eta_2$.  

Choose  $\tilM_n$ to be the smallest integer exceeding $\exp[n  (\frac12 \log(1+S/\sigma_1^2) )-2\gamma)  ]$.  It can be   shown using a  standard change of output measure argument (cf.\ proof of direct part of \cite[Thm.~3.6.2]{Han10}) that with $P_{X^n}$ as the  input distribution in~\eqref{eqn:non_prod_d} and with $\delta$ set to $\gamma/2$,  the decoding error probability  tends to zero.  Choose $M_n$ to be the largest integer smaller than $\exp[n( \frac12 \log(1+S/\sigma_1^2)  - \frac12 \log(1+S/\sigma_2^2)  -4\gamma ) ] = \exp[n(\CG(W;S)-4\gamma)]$ and $\eta_2=\gamma/2$. Thus, $\alpha -\eta_2\ge\frac12 \log(1+S/\sigma_2^2) +\gamma/2$.   With these choices,
\begin{equation}
\bbE\left[\log\frac{W_{\calZ}(\tilZ |\tilX )}{ P_{\tilZ }(\tilZ)}  \right]=\frac{1}{2}\log\Big(1+\frac{S-\gamma/2}{\sigma_2^2}\Big),
\end{equation}
and  from \eqref{eqn:use_iid},
\begin{align}
&\fp_n\le\Pr\bigg( \frac{1}{n}\sum_{i=1}^n\log\frac{W_{\calZ}(\tilZ_i|\tilX_i)}{ P_{\tilZ }(\tilZ_i)}  \ge \frac12 \log\Big(1+\frac{S}{\sigma_2^2} \Big) +\frac{\gamma}{2} \bigg) \nn\\*
&\qquad\qquad\qquad +\exp(-n\eta_2) +\exp(-n\eta_1) . \label{eqn:subs_const}
\end{align}
By the Chernoff bound~\cite[Lem.~6]{Bloch}, the probability in~\eqref{eqn:subs_const} tends to zero exponentially fast. Thus,   $\fp_n$ tends to zero   exponentially fast, proving the lower bound  $C^{(1)}(\bW) \ge \CG(W;S)-4\gamma$. Now take $\gamma\downarrow 0$ to complete the proof of the lower bound.

For the upper bound, $\overC^{(6)}(\bW) \le \CG(W;S)$, we emulate the proof of Theorem~\ref{thm:str_conv_dwtc} with the  (now) unique capacity-achieving output distribution $P_{\barY  \barZ}$ being
\begin{equation}
P_{\barY \barZ}(y,z)=\calN(y;0, S+\sigma_1^2)\calN(z;y, \sigma_2^2-\sigma_1^2). \label{eqn:YbarZbar}
\end{equation}
The derivation up to \eqref{eqn:removed} holds verbatim. So we simply have to check the condition in \eqref{eqn:bound_expect} (with $\CG(W;S)$ in place of $\CDM(W)$) and the behavior of the variance corresponding to~\eqref{eqn:variance}. We first fix an arbitrary sequence $\bx = (x_1, \ldots, x_n)\in\calF_n$ and study the  first two moments of the following information density random variable
\begin{equation}
K^{(n)} (\bx):=\frac{1}{n}\sum_{i=1}^n\log\frac{W_{\calY|\calZ} (Y_i^{(n)}|x_i, Z_i^{(n)})}{P_{\barY |\barZ } (Y_i^{(n)}|Z_i^{(n)})}  .
\end{equation}
We would like to show that $\bbE[K^{(n)}(\bx)]\le \CG(W;S)$ and that the variance of $K^{(n)}(\bx)$ is $O(1/n)$ uniform on $\calF_n$. 
For this task, let $N_j^n$ (for $j=1,2$) be a sequence of i.i.d.\ zero-mean Gaussian random variables with variance $\sigma_j^2$.  Using the constraint $\|\bx\|_2^2\le nS$, the fact that $\sigma_2>\sigma_1$ (as assumed), and the form of the output distributions in~\eqref{eqn:YbarZbar}, it can easily be seen that $K^{(n)}(\bx)$ can be upper bounded as 
\begin{align}
 &K^{(n)}(\bx) \le \CG(W;S) \nn\\*
&\quad+ \frac{\log\rme}{2(1+ \frac{S}{\sigma_1^2}) }\left(\frac{S}{\sigma_1^2} \Big( 1- \frac{\|N_1^n\|_2^2 }{n\sigma_1^2}\Big) +  \frac{2\langle N_1^n,\bx\rangle}{n\sigma_1^2} \right)  \nn\\*
&  \quad  - \frac{\log\rme}{2(1+ \frac{S}{\sigma_2^2}) }\left(\frac{S}{\sigma_2^2} \Big(1- \frac{\|N_2^n\|_2^2 }{n\sigma_2^2}\Big) +  \frac{2\langle N_2^n,\bx\rangle}{n\sigma_2^2} \right) .
\end{align}
Since $N_1^n$ and $N_2^n$ have zero means and covariances $\sigma_1^2\cdot\bI_{n\times n}$ and $\sigma_2^2 \cdot\bI_{n\times n}$ respectively, the expectation  of $K^{(n)}(\bx)$ is bounded above by $\CG(W;S)$. The variance of $K^{(n)}(\bx)$ can be written and bounded as 
\begin{align}
&  \var\big[ K^{(n)}(\bx)\big]    \nn\\*
  &= \var\left[ \frac{1}{n}\log\frac{W_\calY^n(Y^n|\bx)}{P_{\barY}^n(Y^n)}- \frac{1}{n}\log\frac{W_\calZ^n(Z^n|\bx)}{P_{\barZ}^n(Z^n)} \right] \\
& \le 2\var\left[ \frac{1}{n}\log\frac{W_\calY^n(Y^n|\bx)}{P_{\barY}^n(Y^n)}\right]+ 2\var\left[\frac{1}{n}\log\frac{W_\calZ^n(Z^n|\bx)}{P_{\barZ}^n(Z^n)} \right]\\
&\le (2\log^2\rme)\left( \frac{9S^2}{4n(S+\sigma_1^2)} + \frac{\sigma_1^2 S}{n (S +\sigma_1^2)}\right)  \nn\\*
 &\qquad\qquad+ (2\log^2\rme)\left( \frac{9S^2}{4n(S+\sigma_2^2)} + \frac{\sigma_2^2 S}{n (S +\sigma_2^2)}\right) \label{eqn:var_direct}
\end{align}
where \eqref{eqn:var_direct} follows from  direct calculations per \cite[Eq.~(3.7.24)]{Han10} and the fact that $\bx\in\calF_n$. We conclude that uniform over all $\bx\in\calF_n$, the variance of $K^{(n)}(\bx)$ is of the order $O(1/n)$ (depending only on $S,\sigma_1^2,\sigma_2^2$) and hence the Chebyshev argument at the end of the proof of Theorem~\ref{thm:str_conv_dwtc} holds, yielding $\overC^{(6)}(\bW) \le \CG(W;S)$ as desired.
\end{proof}
\subsubsection*{Acknowledgements}  The authors would like to acknowledge several helpful discussions with Mark M.\ Wilde, and especially for bringing our attention to~\cite{morgan} and \cite{renes}. Vincent Tan's  research is supported by NUS startup grants   R-263-000-A98-750/133. Matthieu Bloch's work is supported by NSF with grant CCF1320298, and by ANR with grant 13-BS03-0008.
\bibliographystyle{unsrt}
\bibliography{isitbib}

\end{document}